# Regional study of the Archean to Proterozoic crust at the Sudbury Neutrino Observatory (SNO+), Ontario: Predicting the geoneutrino flux

**Yu Huang**

*Department of Geology, University of Maryland, 237 Regents Drive, Geology Building, College Park, Maryland, 20742 USA (yuhuang@umd.edu)*

**Virginia Strati**

*Dipartimento di Fisica, Università degli Studi di Ferrara, Polo Scientifico e Tecnologico,Via Saragat 1 - I-44122, Ferrara, Italy*

*Istituto Nazionale di Fisica Nucleare, Laboratori Nazionali di Legnaro, Via dell'Università, 2 - I-35020, Legnaro, Padova, Italy*

**Fabio Mantovani**

*Dipartimento di Fisica, Università degli Studi di Ferrara, Polo Scientifico e Tecnologico,Via Saragat 1 - I-44122, Ferrara, Italy*

*Istituto Nazionale di Fisica Nucleare, Sezione di Ferrara, Polo Scientifico e Tecnologico,Via Saragat 1 - I-44122, Ferrara, Italy*

**Steven B. Shirey**

*Department of Terrestrial Magnetism, Carnegie Institution of Washington, Washington, DC 20015, USA*

**Roberta L. Rudnick**

*Department of Geology, University of Maryland, 237 Regents Drive, Geology Building, College Park, Maryland, 20742 USA*

**William F. McDonough**

*Department of Geology, University of Maryland, 237 Regents Drive, Geology Building, College Park, Maryland, 20742 USA*

The SNO+ detector, a new kiloton scale liquid scintillator detector capable of recording geoneutrino events, will define the strength of the Earth's radiogenic heat. The reference Earth model of *Huang et al.* [2013] predicted the crustal geoneutrino signal at SNO+ to be $34.0^{+6.3}_{-5.7}$ TNU (a Terrestrial Neutrino Unit is one geoneutrino event per $10^{32}$ target protons per year), ~60% of which originated from the regional continental crust (the closest six 2° × 2° crustal tiles). A detailed 3-D model of the regional crust, centered at SNO+ and based on compiled geological, geophysical and geochemical information, was used to characterize the physical and chemical attributes of crust and assign uncertainties to its structure. Moho depth in the study area is 42.3 ± 2.8 km and the average upper crust (i.e., the heat producing layer) thickness is 20.3 ± 1.1 km. Monte Carlo simulations were used to predict the U and Th abundances and uncertainties in crustal lithologies and to model the regional crustal geoneutrino signal originating from the at SNO+. The total regional crust contribution of the geoneutrino signal at SNO+ is predicted to be $15.6^{+5.3}_{-3.4}$ TNU, with the Huronian Supergroup near SNO+ contributing $7.3^{+5.0}_{-3.0}$ TNU to this total. The bulk crustal geoneutrino signal at SNO+ is estimated to be $30.7^{+6.0}_{-4.2}$ TNU, which includes the far field crust signal and the regional crust contribution. Finally, without accounting for uncertainties on the signal from continental lithospheric mantle and convecting mantle, the total geoneutrino signal at SNO+ is predicted to be $40^{+6}_{-4}$ TNU.

# 1. Introduction

## 1.1. Motivations

Geoneutrinos, electron antineutrinos generated during beta decays of radioactive nuclides in the Earth, offer a means to determine the concentrations of heat-producing elements (HPEs, namely U, Th, and K), and hence the total radiogenic heat power of the Earth [e.g., *Dye*, 2010; 2012; *Šrámek et al.*, 2012; *Šrámek et al.*, 2013]. A better constraint on the total radiogenic heat power is critical for determining the Earth's heat budget, understanding the power driving plate tectonics, and the thermal and chemical evolution of the planet. Compositional models for the bulk silicate Earth (BSE) differ by a factor of three in U concentration and total radiogenic heat power of the planet [e.g., *McDonough and Sun*, 1995; *Turcotte and Schubert*, 2002; *O'Neill and Palme*, 2008; *Javoy et al.*, 2010]. Geoneutrino data, when available for several sites on the Earth, should be able to define the range of permissible compositional models that describe the BSE.

The physical properties of geoneutrinos have been reviewed in the literature [e.g., *Fiorentini et al.*, 2007; *Dye*, 2012; *Šrámek et al.*, 2012]. The current detection mechanism is the inverse beta reaction, where an anti-neutrino combines with a free proton to produce a positron and a neutron, which are detected by prompt and delayed light flashes recorded in large underground liquid scintillation detectors. This reaction is sensitive to geoneutrinos produced from four beta decays: two each in the $^{238}$U and $^{232}$Th chains; all other geoneutrinos have energies lower than the threshold level (1.806 MeV) that is required to initiate the reaction. Geoneutrinos originating from U and Th decay chains can be distinguished based on their different energy spectra, e.g., only the $^{238}$U chain can produce geoneutrinos with energy >2.25 MeV. KamLAND (Kamioka Liquid scintillator ANtineutrino Detector) in Japan [*Araki et al.*, 2005; *Gando et al.*, 2011; *Gando et al.*, 2013] and Borexino in Italy [*Bellini et al.*, 2010; *Bellini* *et al.*, 2013] are the two detectors that are currently accumulating geoneutrino events, and the experimental results have provided some constraints on the radiogenic heat power from U and Th in the Earth. The SNO+ detector will come on-line in 2014. This kiloton scale detector, a redeployment of the former Sudbury Neutrino Observatory (SNO) at SNOLAB, is located in Ontario, Canada, and will have a significantly higher signal-to-noise ratio for geoneutrino events compared to other operating detectors [*Chen*, 2006]. SNO+ will provide significant new data on the geoneutrino signal mainly originating from U and Th in the surrounding continental crust.

*Huang et al.* [2013] developed a global reference model that predicted the geoneutrino signal from the crust and mantle. Using this model they predicted the geoneutrino flux at KamLAND, Borexino, SNO+, and Hanohano, the latter a proposal for a sea floor detector. The predicted signal from the lithosphere at SNO+ was estimated at $36.7^{+7.5}_{-6.3}$ TNU (a Terrestrial Neutrino Unit is one geoneutrino event per $10^{32}$ target protons per year), of which $34.0^{+6.3}_{-5.7}$ TNU originates from the crust. The far field crust (FFC; defined as the remainder of the crust after removing the closest six 2° × 2° crustal tiles near the detector) contributes $15.1^{+2.8}_{-2.4}$ TNU, and the regional continental crust (closest six tiles) is the dominant geoneutrino source, contributing ~19 TNU. The geoneutrino signal from the mantle at SNO+ is predicted to be between 2 to 16 TNU [*Šrámek et al.*, 2013].

Since SNO+ will accumulate statistically significant amounts of geoneutrino data in the coming years, the calculated signal that is predicted to be derived from the lithosphere can be subtracted from the experimentally determined total geoneutrino signal to estimate the mantle contribution. Such a calculation is key to resolving different BSE compositional models [*Dye*, 2010; *Fogli et al.*, 2012]. It is therefore useful to construct a regional scale reference model based on detailed geological, geochemical and geophysical studies in order to calculate the geoneutrino signal and its uncertainties originating from the local crust. The construction of such a regional reference model is the primary aim of this study.

### 1.2. Constructing the model

We use the published 1:5,000,000 scale Geological Map of North America [*Reed et al.*, 2005] to describe the surface geological characteristics in the six 2°× 2° crustal tiles centered at SNO+ (outlined in Fig. 1), including lithologies, boundaries between different geological terranes/provinces, and their relative proportions. Refraction seismic surveys carried out in this region provide information on the crustal velocity structure and thickness [*Mereu et al.*, 1986; *Epili and Mereu*, 1991; *Percival and West*, 1994; *Musacchio et al.*, 1997; *Winardhi and Mereu*, 1997], whereas reflection seismic surveys and receiver function analysis provide additional constraints on the Moho depth (Fig. 1b) [*Brown et al.*, 1982; *Eaton et al.*, 2006; *Spence et al.*, 2010]. All of the above information is integrated into a 3-D regional crust model. In this model, the layer with P-wave velocities (Vp) between 6.6 and 6.8 km/s is defined as the middle crust, and the underlying layer having Vp between 6.8 and 8.0 km/s is the lower crust. The upper crust is subdivided into seven different lithologic units based on surface exposures (Fig. 2): 1) tonalite and tonalite gneiss of the Wawa-Abitibi sub-provinces, 2) Archean felsic intrusive rocks (granite, granodiorite, etc.), 3) gneissic rocks in the Central Gneiss Belt of the Grenville province, 4) Huronian Supergroup sedimentary to metasedimentary rocks, 5) volcanic/metavolcanic rocks in the Abitibi sub-province, 6) the Southern Province and Sudbury Igneous Complex, and 7) Paleozoic sediments in the southern portion of the study area.

Published databases of litho-geochemical studies performed by the Ontario Geological Survey (OGS) provide high quality U and Th abundance data, determined by ICP-MS or INAA, for most of the major lithologies in the region. The composition of volcanic/metavolcanic rocks in Abitibi is compiled mostly from data in GEOROC. The lake sediments in the Ontario area are assumed to have the same U and Th abundances as the Paleozoic sediments that cover the Great Lakes region. Data from the published literature for the chemical compositions of major lithologies in the study area that are not compiled by GEOROC are also incorporated into the model. Direct samples of the deep crust are limited (lower granulite facies rocks are only exposed in the Kapuskasing Structural Zone, which is at the northwest corner of the six tiles). Therefore, we follow the approach described by *Huang et al.* [2013], which is based on previous studies [e.g., *Christensen and Mooney*, 1995; *Rudnick and Fountain*, 1995] that link the seismic velocity data from refraction seismic surveys with the chemical composition of global amphibolite and granulite facies rocks [*Huang et al.*, 2013], in order to infer the U and Th abundances in the middle and lower crust in the region.

The 3-D model is constructed based on the physical and chemical (U and Th abundances) properties of the regional crust. From these data, the geoneutrino signal at SNO+, as well as the heat production within the study area are calculated, and the associated uncertainties are propagated using Monte Carlo simulations. The geoneutrino signal from the regional model is compared with previous estimates for SNO+ [*Huang et al.*, 2013]. Heat production and heat flow in our model are compared with heat flow measurements in the area [e.g., *Perry et al.*, 2006]. We also evaluate SNO+'s sensitivity to the mantle geoneutrino signal in order to shed light on Earth's chemical composition.

## 2. Geologic framework of the regional crust

The six 2° × 2° crustal tiles centered at SNO+ (approximately 440 km × 460 km), outlined in Fig. 1a, comprise the study area used to construct the 3-D regional reference model and calculate the resulting regional geoneutrino signal. The SNO+ regional crust includes

Precambrian rocks of the southeastern Canadian Shield and Paleozoic sediments of the Great Lakes and Michigan basin. The distribution of the Paleozoic sediments, extending to the Great Lakes Tectonic Zone (GLTZ in Fig. 1a), separates the study area into two distinct portions. The northern portion of the region consists of crystalline rocks of the Neoarchean southeastern Superior Province and the Mesoproterozoic Grenville Province, which borders the southeastern part of the Canadian Shield. The boundary between the Superior Province and Grenville Province is referred to as the Grenville Front Tectonic Zone (GFTZ). The Sudbury Igneous Complex (SIC) and Southern Province are distributed along the GFTZ in the study area (Fig. 1a). The southern portion of the study area is covered by Paleozoic sediments with thicknesses increasing in a southerly direction to the Michigan basin where the total sedimentary cover is up to ~5 km [e.g., *Howell and van der Pluijm*, 1999]. The crystalline basement in the southern region is obscured by this sedimentary cover. In the following sections, the geology of each of these regions is reviewed.

### 2.1. Superior Province

The Superior Province was constructed from fragments of much smaller Archean microcontinents from the Paleoarchean through the Neoarchean and later modified by subsequent collisional events during the Proterozoic. It is one of the world's largest Archean cratons and contains rocks ranging in age from 0.1 to 4.2 Ga. On the basis of lithology, structure, metamorphism, ages, and tectonic events, the Superior Province can be subdivided into several sub-provinces. The Wawa, Abitibi, and Pontiac sub-provinces occur within the six 2° × 2° crustal tiles of our model [e.g., *Card*, 1990; *Percival*, 2007; *Benn and Moyen*, 2008; *Pease et al.*, 2008]. The following geological descriptions are taken from *Card* [1990] and *Percival* [2007].

The Wawa sub-province has an east-west extent of about 600 km, extending east towards the Kapuskasing structure zone (KSZ in Fig. 1a). This sub-province is bounded on the southeast by the Early to Middle Proterozoic Southern Province, on the southwest by the Mid-Continent Rift (Fig. 1b), and on the north by the Quetico and Opatica sub-provinces. The southern extent of the Wawa sub-province is obscured by Lake Superior and Paleozoic sedimentary cover. The eastern Wawa sub-province consists of upper amphibolite facies gneiss that continues gradationally into granulite facies in the KSZ. The Wawa sub-province is composed of Neoarchean juvenile crustal additions of submarine volcanic successions of komatiites, tholeiites and andesites that are intruded by tonalite, granodiorite and granite. Foliated to gneissic tonalitic to granodioritic rocks with zircon U-Pb ages around 2.7 Ga are dominant in the eastern border near the KSZ. The gneiss is cut by younger granodioritic to granitic plutons.

The Abitibi sub-province is the world's largest Archean granite-greenstone belt. The sub-province is bounded on the west by the KSZ and on the east by the GFTZ, a zone of faulting that separates the Superior and Grenville provinces. The southern boundary of the Abitibi sub-province is the GLTZ. The Abitibi region comprises mafic to felsic volcanic/metavolcanic rocks concentrated in the central Abitibi greenstone belt. Tonalite gneiss forms batholithic complexes in and around greenstone belts and is intruded by granitic to granodioritic plutonic rocks.

The small Pontiac sub-province is bounded on the south and east by the GFTZ and on the north by the Abitibi sub-province. It is a Late Archean terrane comprised of lesser exposures of metasediments intruded by tonalite, granodiorite, quartz syenite and granite.

The main lithologies of the Superior Province in the regional model are the tonalite to tonalite gneiss in the Wawa and Abitibi sub-provinces, volcanic/metavolcanic rocks in the Abitibi greenstone belt, and scattered granitic to granodioritic intrusions. The metasedimentary rocks in the Pontiac sub-province are obscured by intrusions.

### 2.2. Grenville Province

The Grenville Province, located to the southeast of the GFTZ, is the primary exposure of the Grenville orogen, which extends from Lake Huron northeastward to the coast of Labrador. In the western Grenville of Ontario, seismic refraction and reflection surveys [e.g., *Mereu et al.*, 1986; *White et al.*, 2000] suggest a tectonic construct involving northwestward stacking of crustal segments to produce the Grenville orogen and an over-thickened crust. The Grenville orogen resulted from ~1100 Ma continent-continent and/or continent-arc collisions and consists of stacked Neoarchean, Paleoproterozoic, and Mesoproterozoic crustal sections [e.g., *Carr et al.*, 2000; *Ludden and Hynes*, 2000]. The Grenville Province comprises a complex assemblage of poly-metamorphosed crustal rocks including plutonic rocks, migmatites, ortho- and para-gneiss, as well as other metasedimentary and metavolcanic rocks. The rocks present in the regional study area range in age from 1.8 to 1.0 Ga. Of the two major belts in the Grenville Province, the Central Gneiss Belt (CGB) and the Central Metasedimentary Belt (CMB) [e.g., *Wynne-Edwards*, 1972], only the former belt is located within the area of interest. The GFTZ is a crustal-scale shear zone marking the northwestern edge of the Grenville orogeny.

The CGB is the oldest part of the Grenville Province, and is dominantly comprised of rocks from the Laurentian craton (pre-1.4 Ga) and younger supra-crustal rocks deposited along the craton margin. The rocks have been subjected to high pressure and high temperature metamorphism during Grenvillian orogenesis. These high-grade metamorphic rocks (e.g., gray gneisses) are exposures of deep sections of Earth's crust to depths of 20 – 30 km. The CMB is a lower grade metasedimentary belt that also corresponds to a lower crustal section. Since the CMB does not fall within the regional study area, we do not consider it further.

Because the seismic surveys coverage in the CGB is insufficient, we cannot identify all the smaller domains within the CGB in the 3-D model. Consequently, we simplify the regional reference model by treating the GFTZ and CGB as a single crustal type and assume that the high-grade gneiss is representative of the dominant lithology.

### 2.3. Southern Province and Sudbury Igneous Complex

The Southern Province comprises a passive margin sedimentary sequence deposited between 2.5 and 2.2 Ga on the southern margin of the Superior Province [e.g., *Long*, 2004; *Long*, 2009]. In the Sudbury region, strata of the Huronian Supergroup crop out extensively in the Southern Province. The main lithologies in the Huronian Supergroup are clastic sedimentary rocks such as sandstones, mudstones, conglomerates, and diamictites (with minor volcanic rocks) that have been metamorphosed at low grades. These sedimentary units crop out along the northern shore of Lake Huron and continue along the GFTZ to the Cobalt Embayment in the northeast. The thickness of the Huronian Supergroup can reach up to ~12 km to the south of the Sudbury Igneous Complex (SIC) [e.g., *Long*, 2009]. Some granitic intrusions with ages 2.1 to 2.3 Ga, including the Skead, Murray and Creighton plutons were emplaced into the Huronian Supergroup.

The Sudbury Structure is a unique geological feature that straddles the boundary between the Superior Province and Southern Province immediately north of the Murray Fault, and about 15 km northwest of the Proterozoic GFTZ. The Sudbury structure is famous for its ore deposits of nickel, copper, cobalt and platinum group elements. The formation of this structure is due to a meteorite impact event 1.85 Ga ago [e.g., *Rousell et al.*, 1997; *Therriault et al.*, 2002]. The three major components of the structure include the Sudbury Basin, the SIC surrounding the basin in the form of an elliptical collar, and an outer zone of Sudbury Breccia [e.g., *Long*, 2009].

The Sudbury Basin is filled with sedimentary rocks of the Whitewater Group, which is approximately 3 km thick and consists of breccias of the Onaping formation, pelagic metasedimentary rocks of the Onwatin formation, and metagraywackes of the Chelmsford formation. The SIC consists of four units: the contact sublayer, norite, quartz gabbro, and granophyre [e.g., *Naldrett and Hewins*, 1984; *Lightfoot et al.*, 1997b]. The latter three units comprise the so-called Main Mass [*Naldrett and Hewins*, 1984]. The relative mass portions of the three units are approximately 40%, 10% and 50%. Footwall rocks consist of Archean granitic and mafic igneous rocks, including granulite facies rocks of the Levack gneiss complex to the north of the Sudbury Structure, and metavolcanic and metasedimentary rocks of the Huronian Supergroup to the south.

### 2.4. Paleozoic sedimentary units

Paleozoic sediments, obscuring the underlying basement, cover approximately 25% of the surface of the study area. The Yavapai, Mazatzal, and Grenville boundaries beneath this Paleozoic sedimentary cover are extrapolated into the crust based on results from *Holm et al.* [2007] and *Van Schmus et al.* [2007]. Only one refraction seismic survey is available for the study area, the GLIMPCE – GLJ line [*Epili and Mereu*, 1991]. For this reason, we simplify the model and interpret the deep crust underlying this Paleozoic cover as extensions of Archean Superior Province and Proterozoic Grenville Province.

### 2.5. Simplified surface geology

The spatial resolution of existing seismic surveys available for determining the crustal structure in our regional model is less than that of the geological map of North America.

Therefore, we have simplified the surface geological map on the basis of characteristic lithologies, metamorphism, tectonic events and U and Th abundances. For the upper crust in the 3-D model, we have identified seven main lithologic units (Fig. 2, Table 5):

1. Tonalite and tonalite gneiss in Wawa and Abitibi sub-provinces;

2. Gneiss in the Central Gneiss Belt (CGB) of the Grenville Province;

3. Granitic to granodioritic intrusions of the Abitibi sub-province;

4. Paleozoic sedimentary cover in Great Lakes area;

5. Volcanic/metavolcanic rocks in the Abitibi greenstone belt;

6. The Sudbury Igneous Complex (SIC);

7. Sedimentary rocks of the Huronian Supergroup.

To construct the 3-D physical model of the study area, the simplified surface geology is combined with information for the vertical crustal structure obtained from refraction and reflection seismic surveys. Receiver function analyses from the Grenville Province provide additional constraints on the crustal thickness in this area [*Eaton et al.*, 2006]. The details of the geophysical inputs into the 3-D model are provided in the next section.

**3. Geophysical 3-D model of the regional crust**

Here we develop the 3-D geophysical model of the main reservoirs of U and Th in the regional crust centered at SNO+, including estimates of the volumes and masses of upper (UC), middle (MC) and lower crust (LC), together with their uncertainties. The seismic velocity data from deep crustal refraction surveys are used to distinguish the three crustal layers [*Mereu et al.*, 1986; *Epili and Mereu*, 1991; *Percival and West*, 1994; *Musacchio et al.*, 1997; *Winardhi and Mereu*, 1997]. In this way, three boundary surfaces are defined in the 3-D model: the top of middle crust (TMC), the top of lower crust (TLC) and the Moho depth (MD). The P-wave velocities 6.6 km/s, 6.8 km/s and 8.0 km/s are adopted as contours to identify these surfaces. The upper crust is further modeled in detail for the seven dominant lithologic units as defined in the previous section by combining the simplified geological map and vertical crustal cross sections.

### 3.1. Geophysical model of boundary surfaces

The geophysical inputs used for estimating the depth of the TMC, TLC and MD come from seismic surveys and receiver function analyses. To construct this 3-D model, five main reflection and refraction seismic experiments, performed over the last 30 years (Table 1), were used over an extended area that includes the six tiles (Fig 1b). Some interpreted crustal cross sections based upon these seismic experiments (Table 2) were used to define the contacts between the seven lithologic units in the upper crust [*Geis et al.*, 1990; *Percival and Peterman*, 1994; *Adam et al.*, 2000; *Easton*, 2000; *Ludden and Hynes*, 2000; *White et al.*, 2000].

The Lithoprobe Abitibi-Grenville Seismic Refraction Experiment, which began in 1992, aimed to explore the main tectonic features of the Grenville and Superior provinces in the southeastern Canadian Shield [*Winardhi and Mereu*, 1997]. This project includes the acquisition of about 1250 km of seismic profiles along four long refraction lines. The profiles XY and AB, forming a cross-arm array centered a few kilometers away from SNO+, are particularly important for modeling the crustal structure directly beneath the detector (Fig. 1b). Profile EF images seismic discontinuities of the deep crustal layers beneath the Abitibi greenstone belt. The Central Metasedimentary Belt (CMB), the Central Gneiss Belt (CGB), the Grenville Front Tectonic Zone (GFTZ), the Pontiac sub-province and the Abitibi greenstone belt are investigated by the long refraction line MG. The five refraction profiles across the Kapuskasing Structural Zone ["PW" in Fig. 1b; *Percival and West*, 1994] and the GLJ line in Lake Huron [*Epili and*

*Mereu*, 1991] are used for constraining the crustal structure of the northern and southern region, respectively.

To increase the quality of the 3D model around the borders of the regional crust, we used data from four other seismic experiments that occur outside of the study area. The Ontario-New York-New England (O-NYNEX) refraction profile SP in the Appalachian Province [*Musacchio et al.*, 1997] and four refraction surveys made by the Canadian Consortium for Crustal Reconnaissance (COCRUST) [*Mereu et al.*, 1986] across the Ottawa Graben are used to constrain the eastern boundary of our study area. Three reflection lines, GLA [*Spence et al.*, 2010], MIC1 and MIC2 [*Brown et al.*, 1982], located in the marginal western and southern area (Lake Superior and Michigan Basin) provide data on Moho depth in the southwestern region of the study area, but no refraction studies are available for this region.

The crustal thickness in the Grenville Province is obtained from receiver function analyses of 537 seismic events registered by 32 broadband seismic stations (blue triangles in Fig. 1b) [*Eaton et al.*, 2006]. The average uncertainty for the Moho depth obtained by the receiver function technique is ± 0.8 km, which is negligible and is not included in our model.

The Moho depth at the 32 teleseismic stations are combined with 343 and 22 data points digitized on 15 refraction and 3 reflection sections, respectively, in order to determine the depth of the different crustal layers. The average interval of digitalization along the 4552 km of refraction and reflection lines is 12.5 km. For refraction tomography, 343 depth points were digitized following the velocity contours of 6.6 and 6.8 km/s, which define the TMC and TLC surfaces. In Table 3 we report the descriptive statistics of depth-controlling points along each of the three boundary surfaces.

Depth maps for TLC, TMC and MD surfaces are obtained by applying a geostatistical estimator (ordinary kriging) that infers the depths of these surfaces for locations without direct observations by interpolating between depth-controlling data points. The main advantages of this method are that it takes into account the spatial continuity of the depths and it provides the uncertainties of the estimated depths of three surfaces for each grid cell. The continuous depth maps for the TMC, TLC and MD surfaces (Fig. 3a, c, e) are obtained using this method. We also report the Normalized Estimation Errors (NEE) maps (Fig. 3b, d, f) that provide estimated uncertainties in terms of variance, normalized with respect to depth and expressed in percentage.

The average Moho depth in the six tiles is $42.3 \pm 2.6$ km, while the average TMC and TLC depths are $20.3 \pm 1.1$ km and $26.7 \pm 1.5$ km, respectively. The uncertainties associated with the average depths of TMC and TLC are correlated, and their sum corresponds to average uncertainty in the MD. Table 4 reports the average thickness and volumes of upper, middle and lower crust within the six tiles. The regional crustal model yields a very similar estimate of average crustal thickness ($42.3 \pm 2.6$ km) as the global crustal reference model of *Huang et al.* [2013] for the same region ($42.6 \pm 2.8$ km). These two crustal thickness estimates are larger than the estimate provided in the new global crustal model CRUST 1.0, which yields an average crustal thickness of only 39.2 km for the study area, without uncertainty [*Laske et al.*, 2013]. The regional crustal model is based on *in situ* seismic surveys, which were carried out to understand the deep crustal structure, and thus provide a more accurate and precise evaluation of the crustal thickness and its uncertainty in this region than any global-scale crustal model. Another obvious difference between the regional crustal model and previous global models is the relative thickness of middle crust, which is very thin in the new 3-D regional model. This occurs because of our selection of the Vp interval from 6.6 to 6.8 km/s as the velocity range of the middle crust, which results in the thin middle crust.

The densities of the middle and lower regional crust are estimated using the functional relationship between Vp and density provided by *Christensen and Mooney* [1995] (see their Table 8). Since the depth to center of the middle crust in the study area is about 25 km, we use the average densities obtained using the functions $\rho = a + b/Vp$ (where *a* and *b* are the empirically determined coefficients linking Vp and density) applied to 20 km, and 30 km depths, respectively, with average middle crust Vp as $6.7 \pm 0.1$ km/s. The center of the lower crust in the study area is about 35 km, and most of the layer has Vp of $7.0 \pm 0.2$ km/s. Thus, we use the functions for 30 km and 40 km depth. The average density of middle crust and lower crust are estimated to be $2.96 \pm 0.03$ and $3.08 \pm 0.06$ g/cm$^3$, respectively. The model of *Huang et al.* [2013] adopted the density information provided in CRUST 2.0 [*Bassin et al.*, 2000], and the updated CRUST 1.0 [*Laske et al.*, 2013] provides an even lower density crust (Table 4). The average densities for this region in the previous two global crust models are both lower than what we obtain here. This is perhaps due to the fact that the middle crust in the regional model is deeper and thinner, and the lower crust is thicker than that of CRUST 1.0. The density of lower crust in *Huang et al.* [2013] falls within 1-sigma uncertainty of that in our regional model.

**3.2. Geophysical model of regional upper crust**

As described previously, the geological map of the regional upper crust has been simplified into seven dominant lithologic units (Fig. 2). The exposed contacts between the seven lithologic units are combined with interpreted crustal cross sections from the seismic profiles to construct the physical upper crust model. The physical properties (thickness, volume, density and mass) of each lithologic unit are reported in Table 5.

The tonalite or tonalite gneiss in the Wawa and Abitibi sub-provinces is the dominant lithology within the study area, accounting for 60% of the total volume of the regional upper crust. The top of this unit corresponds to the bottom of some minor lithologic units, or a flat Earth's surface, while its bottom is the TMC surface underneath the Superior Province, as revealed by interpreted crustal cross sections. Gneissic rocks in the CGB of the Grenville Province is the next most voluminous lithology, encompassing ~30% of the total volume of the regional upper crust; its surface exposure is not obscured by any other lithologies except for a thin veneer of Paleozoic sedimentary rocks in the southern portion of the study area (Fig. 4a and e). All other five lithologic units account for the remaining 10% of upper crustal volume. The thicknesses of these remaining lithologic units are constrained from interpreted crustal cross sections based on reflection seismic surveys. The thickness of the Paleozoic sedimentary cover in the Great Lakes region is further constrained by the contours of thicknesses of sedimentary rocks in the Michigan Basin [*Howell and van der Pluijm*, 1999]. South of the SIC, the thickness of the Huronian Supergroup can reach up to 12 km [e.g., *Long*, 2009] and, thus, for the purpose of better constraining its thickness, some virtual points with depths of 12 km are added to the 3-D model. The thickness of several regionally distributed granitic to granodioritic intrusions were estimated using reflection seismic surveys. The structure of the SIC has been extensively explored [e.g., *Milkereit et al.*, 1994; *Boerner et al.*, 2000; *Snyder et al.*, 2002] and existing seismic surveys are sufficient to constrain the SIC precisely. Some extrapolations regarding the thickness of the volcanic to metavolcanic rocks in the Abitibi greenstone belt were made based on existing geological cross-sections in order to estimate their thickness in areas lacking direct observations.

The density of each upper crustal lithologic unit is obtained from the literature. *Fountain et al.* [1990], *Salisbury and Fountain* [1994] and *Fountain and Salisbury* [1996] provide laboratory density measurements for a variety of samples from the Canadian Shield. The samples are here reclassified into tonalite-tonalite gneiss, granite-granodiorite and greenstone belt volcanic rocks and the average and standard deviation of the density are used for these lithologic units. The CGB gneissic rocks are assumed to have similar density to tonalite gneiss of comparable metamorphic grade. *Hinze et al.* [1978] provide drill core density information for sedimentary rocks in the Michigan Basin and we adopt this density for the Paleozoic sedimentary rocks. Densities of sedimentary rocks in the Huronian Supergroup are obtained from Ontario Geological Survey published preliminary map 2297. The density of the SIC is obtained from drill core information published by *Snyder et al.* [2002] and *Milkereit et al.* [1994]. Since 90% of the regional upper crust has a density of $2.73 \pm 0.08$ g/cm$^3$, the average upper crust is assumed to have the same density. With the volume and density information, the mass of each lithologic unit was calculated (Table 5).

**3.3. Cross-checking the 3-D model**

Six schematic E-W and N-S cross-sections of the region show the results of the numerical 3-D model (Fig. 4), with the correct positions and shapes of major tectonic features, such as the GFTZ in cross sections A-A', C-C' and D-D', and the SIC in cross sections A-A' and B-B'. The thickness of each minor lithologic units in the upper crust agrees with the interpreted cross-sections based on seismic surveys.

**4. Chemical composition of the crust near SNO+**

The abundances of U and Th in the seven dominant, upper crustal lithologic units were evaluated based on analyses of representative outcrop samples. The OGS has published litho-geochemical databases that provide high quality U and Th abundance data (obtained mostly by inductively coupled plasma mass spectrometry -- ICP-MS, and sometimes by instrumental neutron activation analysis -- INAA) for some of the dominant lithologic units, such as the SIC, the Huronian Supergroup sedimentary rocks, granitic to granodioritic intrusions, and tonalite/tonalite gneiss. In addition, new unpublished compositional data for glacial tillites in the Huronian Supergroup were provided by Dr. Richard Gaschnig [*Gaschnig et al.*, 2014]. Data for the Abitibi greenstone belt volcanic rocks come from GEOROC, a web-based geochemical database, which contains a compilation of compositional data for volcanic to metavolcanic rocks from the belt. This compilation has also been supplemented with OGS data. For the CGB of the Grenville Province, there is limited U data, but abundant ICP-MS Th data provided by Dr. Trond Slagstad (NGU, Norway). The lake sediments in Ontario are assumed to be sourced by Paleozoic sedimentary rocks. We use the composition of these lake sediments to represent the composition of the Paleozoic sedimentary rocks.

The quality of the dataset is improved by excluding all U and Th data determined by XRF, which is less sensitive and accurate compared to ICP-MS or INAA data. Some lithologic units are further subdivided into end members according to lithology/composition, such as Abitibi volcanic rock, the SIC, and felsic intrusions. The volcanic rocks of the Abitibi greenstone belt are a mixture of felsic, intermediate, and mafic rocks with relative mass proportions of 5%, 40% and 55%, respectively [*Card*, 1990]. The Main Mass of the SIC is composed of norite (40%), quartz gabbro (10%), and granophyre (50%) [e.g., *Lightfoot et al.*, 1997a]. Granite (60%) and granodiorite (40%) are the two dominant types of intrusive rocks. These proportions are obtained from the geological map by comparing the surface exposure areas of the two rock types in the red unit. For these three lithologic units, the weighted average composition is obtained using Monte Carlo simulations [*Huang et al.*, 2013] to predict the geoneutrino signals from these units.

After constructing the databases, average U and Th abundances for each of the lithologic units can be estimated, along with uncertainties. It has long been known that the frequency distributions of the abundances of highly incompatible elements, such as U and Th, in typical crustal reservoirs are strongly skewed and fit a log-normal distribution, rather than a Gaussian distribution [e.g., *Ahrens*, 1954; *Huang et al.*, 2013]. To reduce the influence of rare enriched or depleted samples on the log-normal average, we apply a 1.15-sigma filter that removes about 25% of the data. Then the median value of the distribution is used to represent the central tendency of the distribution, and 1-sigma uncertainty covers 68.3% of the remaining data population. The abundances of U and Th in each lithologic unit in the regional upper crust are reported in Table 6.

The U abundances in the tonalite and CGB gneiss, which together account for 90% of the upper crustal volume in the 3-D model, are $0.7^{+0.5}_{-0.3}$ ppm and $2.6 \pm 0.4$ ppm, and the Th abundances of these two units are $3.1^{+2.3}_{-1.3}$ ppm and $5.1^{+6.0}_{-2.8}$ ppm, respectively. There are only five high quality measurements of U abundances in the CGB gneiss, which is the key limitation of the statistics. This is also the reason why the U abundance in the CGB gneiss has a symmetric uncertainty. The Wawa-Abitibi volcanic rocks are depleted in HPEs compared to other lithologic units in the study area. The Huronian Supergroup metasedimentary rocks are the closest unit to the SNO+ detector and are enriched in U $4.2^{+2.9}_{-1.7}$ ppm and Th $11.1^{+8.2}_{-4.8}$ ppm relative to the greenstone belt rocks; therefore, this minor lithologic unit, due to its proximity to the detector, could be the dominant source of the geoneutrino signal at SNO+.

The composition of the middle and lower crust in the 3-D model is inferred from seismic velocity data from the refraction seismic surveys, using the same approach as that adopted by *Huang et al.* [2013]. The seismic velocity in the middle crust is 6.7 ± 0.1 (1-sigma) km/s, and for the top 70% of the lower crust is 7.0 ± 0.2 (1-sigma) km/s while the lower 30% has velocity >7.2 km/s.

Compiled laboratory ultrasonic velocity measurements of deep crustal samples taken at room temperature and various confining pressure are used to estimate the composition of the deep crust. At 600 MPa, felsic amphibolite-facies rocks have an average Vp of 6.34 ± 0.16 (1-sigma) km/s, while mafic amphibolite-facies rocks have a Vp of 6.98 ± 0.20 km/s. Felsic granulite-facies rocks have an average Vp of 6.52 ± 0.19 km/s, while mafic granulites have an average Vp of 7.21 ± 0.20 km/s [*Huang et al.*, 2013]. The temperature and pressure correction derivatives are $-4 \times 10^{-4}$ km s$^{-1}$ °C$^{-1}$ and $2 \times 10^{-4}$ km s$^{-1}$ MPa$^{-1}$ [e.g., *Kern*, 1982; *Jackson*, 1991; *Christensen and Mooney*, 1995; *Rudnick and Fountain*, 1995]. Assuming a typical conductive geotherm [*Pollack and Chapman*, 1977] corresponding to a surface heat flow of 50 mW m$^{-2}$ (the observed average surface heat flow in the Sudbury region [*Perry et al.*, 2009]), the temperature in the crust increases linearly to 600°C at 50 km depth. The average depth of middle crust in the regional model is about 25 km, corresponding to a pressure of ~700 MPa and a temperature of ~300°C. Therefore, the pressure-temperature corrected felsic amphibolites have a Vp of 6.25 ± 0.16 km/s and mafic amphibolites have a Vp of 6.89 ± 0.20 km/s. For the lower crust, the pressure is ~1000 MPa and the temperature is ~400°C; therefore, the corrected felsic granulites have Vp of 6.45 ± 0.19 km/s and mafic ones have 7.14 ± 0.20 km/s. Given the Vp in the middle and top layer of lower crust is set to be 6.7 ± 0.1 km/s and 7.0 ± 0.2 km/s in the model, the fractions of felsic end member in the two reservoirs are 0.3 ± 0.4 and 0.2 ± 0.4, which are combined with the compositions of felsic and mafic end members from *Huang et al.* [2013] to calculate the U and Th abundances in the deep crust.

The abundances of U and Th in any single lithologic unit (except the CGB gneiss, for which only five U analyses are available) in the regional model are significantly correlated. Table 6 reports the correlation between logarithmic values of U and Th abundance in each reservoir. This correlation introduces a non-negligible effect on the uncertainties of the geoneutrino signal and radiogenic heat power. Previous reference models (global scale or regional scale) for the geoneutrino signal [e.g., *Mantovani et al.*, 2004; *Dye*, 2010; *Coltorti et al.*, 2011; *Huang et al.*, 2013] ignored the correlation between U and Th abundances. Here we use Monte Carlo simulations to incorporate fully the variations in U and Th abundances and the influence of the correlation between them in estimating uncertainties.

## 5. Geoneutrino signal

In order to predict the geoneutrino signal and assign precise oscillation parameter values, the 3-D regional crustal model was divided into cells of 1 km × 1 km × 0.1 km dimension, generating a grid with about $9 \times 10^8$ cells. In each of these cells spatial, geophysical, and geochemical data were assigned.

Using the abundance and distribution of U and Th in the 3-D regional crustal model given above, the regional contribution to the geoneutrino signal at SNO+ is calculated by summing the geoneutrino signal produced by U and Th in each of the $9 \times 10^8$ cells. The geoneutrino production rates from U and Th are $7.41 \times 10^7$ and $1.62 \times 10^7$ kg$^{-1}$ s$^{-1}$, respectively [e.g., *Fiorentini et al.*, 2007]. The oscillated geoneutrino flux at SNO+ from the center of each cell is calculated by taking into account three-flavor survival probability $P_{ee}$ and the geoneutrino energy spectrum [*Fiorentini et al.*, 2012; *Fogli et al.*, 2012]. The 1-sigma uncertainty associated with the survival probability introduces an uncertainty of a few percent to the geoneutrino signal, which is negligible when compared to the other uncertainties (e.g., U abundance). The conversion factor from geoneutrino flux to reported geoneutrino signal is calculated by integrating the geoneutrino energy spectrum and inverse beta decay reaction cross section, and is $12.8 \times 10^{-6}$ (TNU cm$^2$ s) and $4.04 \times 10^{-6}$ (TNU cm$^2$ s) for U and Th, respectively [*Fiorentini et al.*, 2007; *Fiorentini et al.*, 2012].

Monte Carlo simulation, as described by *Huang et al.* [2013], is employed to predict the uncertainty of the geoneutrino signal at SNO+. The uncertainties for thickness and density of each lithlogic unit, correlated abundances of U and Th, and fractions of felsic end members in the middle and lower crust are expressed by distributions of generated random number matrices. These input matrices are propagated through the geoneutrino signal calculation to obtain the distribution of the geoneutrino signal. The median value is chosen to describe the central tendency of the skewed output distribution, which is due to the propagation of log-normal distributions of U and Th abundances.

The predicted geoneutrino signals from U and Th and associated uncertainties (1-sigma) are reported in Table 7 for the seven dominant lithologic units in the upper crust, the middle and the lower crust. The uncertainties on the geoneutrino signals are apparently dominated by the uncertainties in U and Th abundances. For the upper crust, the Huronian Supergroup is the dominant geoneutrino source ($7.3^{+5.0}_{-3.0}$ TNU; 55% of the total contribution of the upper crust) because of its proximity to SNO+ and its high U and Th abundances. The tonalite/tonalite gneiss in the Wawa-Abitibi sub-provinces and CGB gneissic rocks contribute comparable geoneutrino signals (~2 TNU) at SNO+. The felsic intrusions and the SIC both contribute less than 1 TNU. The other two lithologic units (volcanic rocks in the Abitibi greenstone belt and Paleozoic sedimentary rocks in Great Lakes area) add negligible contributions. The total geoneutrino signal at SNO+ from the regional crust is predicted to be $15.6^{+5.3}_{-3.4}$ TNU, 85% of which originates from the upper crust.

It is of note that the Elliot Lake uranium deposit occurs within the study area (Fig. 1a). However, this deposit does not produce a significant geoneutrino signal at SNO+. Uranium mining in the area of Elliot Lake, Ontario was initiated in the mid-1950s, but operations have now ceased. The U abundances in the ores ranged between 100-1000 ppm. These inactive uranium mines are ~100 km to the west of the SNO+ detector. Assuming an exceptionally generous total volume of all the mines in Elliot Lake reached 5 km × 5 km × 5 km (n.b., 5 km is projected to 1 mm in the 1:5M geology map used to construct the regional crust model) and the density of the ore is 3000 kg/m$^3$, the total mass of U is $3.8 \times 10^{10}$ to $3.8 \times 10^{11}$ kg. Given the geoneutrino activity of U is $7.64 \times 10^7$ kg$^{-1}$ s$^{-1}$ [*Fiorentini et al.*, 2007], the geoneutrino luminosity of these deposits in Elliot Lake is $2.9 \times 10^{18}$ to $2.9 \times 10^{19}$ s$^{-1}$, and the corresponding geoneutrino flux at the SNO+ is only $2.3 \times 10^3$ to $2.3 \times 10^4$ cm$^{-2}$ s$^{-1}$, which translates into a geoneutrino signal of 0.02 to 0.2 TNU. Comparing to the total U signal at SNO+ originating from the regional crust (Table 7), the Elliot Lake U deposit can contribute only 0.2% to 2%, which is negligible given that the uncertainty on the geoneutrino signal is near 30%. For this reason, the Elliot Lake uranium deposit is not included in our regional crust model.

# 6. Discussion

### 6.1. Heat production

The estimates of U and Th abundances in different upper crustal lithologic units are based on measurements of outcrops of representative rock types. The heat flow measurements carried out in the Canadian Shield [e.g., *Perry et al.*, 2006] provide a test of the 3-D regional crustal model.

We calculate the heat production in each of the units in the regional crust model (Table 7). The heat production and its uncertainty strictly depend on the U and Th abundances. We have not included K in our regional model, as geoneutrinos generated during its decay cannot be detected using current technology. We therefore assume that the K/U in the crust is about $10^4$ [e.g., *Wasserburg et al.*, 1964; *Rudnick and Gao*, 2003], and therefore the heat production of K is 1/3 of that of U.

For the Superior Province, the tonalite/tonalite gneiss in the Wawa and Abitibi sub-provinces dominates the heat production in the upper crust. The calculated heat production of the upper crust of the Superior Province is 0.62 $\mu W\ m^{-3}$ (with contributions from U, Th and K of 0.25, 0.29 and 0.08 $\mu W\ m^{-3}$, respectively), while that of the middle crust is 0.54 $\mu W\ m^{-3}$ and the lower crust is 0.18 $\mu W\ m^{-3}$. Given a thickness of upper, middle and lower crust in the Superior Province of 20.3, 6.4 and 15.6 km, respectively, then the heat flow contribution from the bulk crust in the Superior Province (Wawa and Abitibi sub-provinces region) is estimated at 19 mW $m^{-2}$. If the mantle heat flow in this area is between 12-18 mW $m^{-2}$ [e.g., *Perry et al.*, 2006], the average surface heat flow in the Superior Province is between 31-37 mW $m^{-2}$, which is lower than the measured average heat flow of 40.9 ± 0.9 mW $m^{-2}$ [*Perry et al.*, 2006, where the uncertainty is expressed as the standard error of the mean and is likely to be an underestimate of the true uncertainty]. Given that the uncertainty on the estimate of crustal heat flow contribution from our model is greater than 25% (± 5 mW $m^{-2}$) or even higher considering the variations of U and Th abundances, the heat flow measurements and the 3-D regional model agree.

The upper crust of the CGB of the Grenville Province is relatively enriched in U and Th with a heat production of 1.29 μW m$^{-3}$ compared to the Superior Province. Thus, given the thickness of upper, middle and lower crust in Grenville Province is 14.5, 6.4 and 15.6 km, the crustal contribution to surface heat flow is 25 mW m$^{-2}$. The Grenville Province also has an average surface heat flow of 41 mW m$^{-2}$ [*Mareschal and Jaupart*, 2004]. Assuming the mantle heat flow is 12-18 mW m$^{-2}$, the total surface heat flow is predicted to be 37-43 mW m$^{-2}$. Thus, the 3-D regional crustal model predicts a surface heat flow that agrees with surface measurements for the Grenville Province.

Heat flow measurements made near the SIC yield an average of 53 mW m$^{-2}$ [*Perry et al.*, 2009]. The locally enhanced heat flow is interpreted as being due to the thick Huronian Supergroup and granitic intrusions that are present around the SIC. The 3-D model shows a maximum thickness of Huronian Supergroup to the south of the SIC of 12 km, which can itself contribute a heat flow of 27 mW m$^{-2}$. If the rest of the upper crust is high-grade gneissic rock such as tonalite gneiss [*Lightfoot et al.*, 1997a], this unit can generate 4 mW m$^{-2}$. With 6 mW m$^{-2}$ from deep crust and 12-18 mW m$^{-2}$ mantle flow, the surface heat flow near the SIC is calculated to be 49-55 mW m$^{-2}$ for the 3-D model presented here. This range agrees with the surface heat flow measurements. However, the 3-D regional model does not have sufficiently high resolution to enable one to determine the variation of surface heat flow in the area near the SIC due to the lack of dense seismic survey coverage.

**6.2. Mantle geoneutrino signal**

The motivation for undertaking this regional crustal study in the Sudbury area is, in part, to determine whether the SNO+ detector, on its own, has the sensitivity to discriminate the mantle geoneutrino signal, which can be obtained by subtracting the crustal signal from future experimentally measured signal ($S_{tot,\ meas}$):

$$S_{mantle} = S_{tot,meas} - S_{FFC} - S_{LOC.}$$

The crustal signal is divided into two components FFC (Far Field Crust) and LOC (LOcal Crust). *Huang et al.* [2013] predicted the FFC signal at SNO+ to be $15.1^{+2.8}_{-2.4}$ TNU and the LOC to be $18.9^{+3.5}_{-3.3}$ TNU.

Our regional crustal study predicts an updated LOC signal of $15.6^{+5.3}_{-3.4}$ TNU, which is within 1-sigma uncertainty of the prediction obtained in the global reference model. The uncertainty for the LOC signal from the regional crustal study is larger than the uncertainty from the global reference model. The main reason for the larger uncertainty is the large variations in U and Th abundances in each of the dominant upper crustal lithologic units, especially the Huronian Supergroup (Table 6). Our model yields a larger uncertainly on U and Th abundances in the regional crust than global model, since the grid on the global model was so coarse that it could not account for regional variability. The only exception is the SIC, for which only a single dataset from an OGS Open File Report 5959 [*Lightfoot et al.*, 1997c] is used to estimate the U and Th abundances. Considering the much larger surface exposures of other lithologic units, the variations of rock type and chemical composition become larger.

The bulk crustal geoneutrino signal at SNO+ is estimated to be $30.7^{+6.0}_{-4.2}$ TNU, by summing the FFC signal [*Huang et al.*, 2013] with LOC signal from this study. Since the FFC and LOC geoneutrino signals at SNO+ are independently estimated, their individual uncertainties should be summed up in a quadratic approach, rather than linear. In this way, this regional crustal study for Sudbury slightly reduces the uncertainty on the predicted bulk crustal geoneutrino signal compared to the prediction based on global reference model, which is $34.0^{+6.3}_{-5.7}$

TNU [*Huang et al.*, 2013]. The continental lithospheric mantle is predicted to produce about 2 TNU at SNO+ [*Huang et al.*, 2013]. Assuming a BSE compositional model of *McDonough and Sun* [1995], the predicted mantle signal at SNO+ is 7 TNU [*Šrámek et al.*, 2013]. Without taking into account the uncertainties on the signal from continental lithospheric mantle and convecting mantle, the total geoneutrino signal at SNO+ is predicted to be $40^{+6}_{-4}$ TNU.

For the sake of simplicity, we adopt an uncertainty on the bulk crustal signal at SNO+ as 5.1 TNU in the discussion below. Following the equations defined by *Dye* [2010], the mantle signal determination at SNO+ has a systematic uncertainty of 2.1 TNU, assuming the systematic uncertainty for nuclear reactor events is 0.9 TNU and for detection exposure determination is 1.9 TNU. Counting uncertainty for geoneutrino detection decreases with accumulation of geoneutrino events following Poisson's law. Assuming SNO+ can "see" 30 geoneutrino events per year, the counting uncertainty on the geoneutrino detection rate drops to 11% of 40 TNU after three years. By summing the counting and systematic uncertainties for the geoneutrino event rate at SNO+, the combined uncertainty becomes 4.7 TNU. Therefore, the mantle signal determination at SNO+ has a total uncertainty of 6.9 TNU after three years of full operation and after subtracting the crustal signal from the measured total signal.

*Šrámek et al.* [2013] predicted the mantle geoneutrino signal at SNO+ using various BSE compositional models and mantle structure to range between 2 to 16 TNU. To resolve these different mantle signals, the uncertainty for determining the mantle signal should be, at most, 7 TNU, which is the same as SNO+'s sensitivity to the mantle signal as described above. Unfortunately, the current constraints on the abundances and distributions of U and Th in the regional crust, especially in the Huronian Supergroup, are not good enough to make the goal of resolving the signal from the mantle feasible. There are three possible ways to advance this situation: 1) improve our knowledge of the distribution of the Huronian Supergroup by undertaking higher density seismic surveys, as well as performing a systematic study of the U and Th abundances in different representative metasedimentary/sedimentary rocks; 2) combing the experimental results at the three operating detectors to determine the mantle signal, rather than relying on any single detector; 3) conducting a geoneutrino experiment, such as the proposed Hanohano detector [*Learned et al.*, 2008], which is carried out in the oceans in order to minimize the signal from continental crust.

Performing high-density seismic surveys within the Huronian Supergroup would allow better estimation of its physical structure with smaller uncertainty (~10%). The geophysical uncertainty will contribute ~±1 TNU on the geoneutrino signal originating from the Huronian Supergroup. The relatively large variations of U and Th abundances in the Huronian Supergroup are the dominant sources of uncertainty on the geoneutrino signal from this source. Ideally, detailed geochemical mapping of U and Th abundance variations in the area will improve the geoneutrino signal prediction significantly. If the geochemical uncertainties on the Huronian Supergroup (e.g., U and Th abundances) can be reduced to 10 to 20%, they would contribute ~±1 to ±2 TNU uncertainty on the geoneutrino signal. Combining the geophysical and geochemical uncertainty contributions, higher resolution studies of the Huronian Supergroup are likely to reduce the uncertainty on the geoneutrino signal from this source to ~±2 TNU, smaller than the $^{+5.0}_{-3.0}$ TNU obtained in this study. Given that this task is feasible in the near future, the uncertainty on the LOC geoneutrino signal at SNO+ could be reduced to ~±2.5 TNU, and the bulk crust signal would have an uncertainty ~±3 TNU. Following this approach, the extracted mantle signal from future experimental results at SNO+ will have an uncertainty ~±5.6 TNU, which would allow better resolution of the various BSE compositional models.

**7. Conclusions**

We have constructed a 3-D regional crustal reference model aimed at predicting the geoneutrino signal at SNO+, a third geoneutrino detector located in Sudbury, Ontario, Canada, which is scheduled to come on-line in 2014. The uncertainty of the predicted geoneutrino signal is estimated through Monte Carlo simulation, and stems mainly from the large uncertainties on U and Th abundances in the upper crust. The main results of this study are as follows:

1. Surface geology, refraction and reflection seismic surveys, receiver function analyses, and interpreted vertical crustal cross sections in the Sudbury region are integrated into a 3-D model for the six 2° × 2° tiles centered at SNO+. The average thickness of the regional crust is estimated to be 42.3 ± 2.6 km, which is in good agreement with the results obtained in the previous global crustal reference model [*Huang et al.*, 2013]. The thickness of upper crust reaches 20.3 ± 1.1 km, which accounts for about half of the bulk regional crust.

2. The upper crust is subdivided into seven dominant lithological units in order to improve the resolution in predicting the geoneutrino signal. The tonalite/tonalite gneiss in the Wawa and Abitibi sub-provinces is the dominant rock type (60% of upper crust), and has $0.7^{+0.5}_{-0.3}$ ppm U and $3.1^{+2.3}_{-1.3}$ ppm Th. The low proportion of HPEs relative to global average upper continental crust reflects the fact that the Superior Province is an Archean craton with intrinsically low heat production, consistent with other Archean cratons globally. The high-grade gnessic rock, with 2.6 ± 0.4 ppm U and $5.1^{+6.0}_{-2.8}$ ppm Th, in the Central Gneiss Belt (CGB) of Grenville Province is the second dominant rock type in the 3-D model (30% of upper crust). Huronian Supergroup metasedimentary rocks are the closest unit to the SNO+ detector and are enriched in U $4.2^{+2.9}_{-1.7}$ µg/g and Th $11.1^{+8.2}_{-4.8}$ µg/g relative to the other units. All of the uncertainties are propagated through Monte Carlo simulations to the geoneutrino signal prediction.

3. The total regional crust contribution of the geoneutrino signal at SNO+ is predicted to be $15.6^{+5.3}_{-3.4}$ TNU. This signal is somewhat lower than the prediction made using the global reference model [*Huang et al.*, 2013], which is $18.9^{+3.5}_{-3.3}$ TNU. This difference is likely to be due to the lower HPEs in Archean to Proterzoic rocks of the Canadian Shield relative to the global average bulk upper continental crust. Considering the uncertainty, the two predictions of geoneutrino signal are consistent with each other. The Huronian Supergroup is predicted to be the dominant source of the geoneutrino signal and its uncertainty at SNO+, and variation in U and Th abundances within the Huronian Supergroup is the primary source of the large uncertainty on the predicted geoneutrino signal.

4. Assuming that the continental lithospheric mantle and convecting mantle together contribute a 9 TNU signal to SNO+, the total geoneutrino signal at SNO+ is estimated to be $40^{+6}_{-4}$ TNU.

5. The large uncertainty in the crustal geoneutrino signal indicates that SNO+, on its own, is unlikely to provide constraints on the mantle geoneutrino signal, and therefore to the debate regarding BSE compositional models. Several future approaches to constrain mantle geoneutrino signal are: improve the 3-D regional model for predicting the regional crustal contribution by increasing the resolution of the distribution of the Huronian Supergroup using a higher density of local seismic surveys and geochemical analyses; combining experimental results at the three operating detectors to place better constraints on the mantle signal; and conducting geoneutrino detection in the oceans, such as the proposed Hanohano detector, in order to minimize the signal from continental crust.

**Acknowledgements:**

We are grateful many people for providing help and sharing data for geoneutrino prediction: John Ayer for leading a field trip to Sudbury, Mike Easton and John for sharing their chemical composition data for the Canadian Shield, Richard Gaschnig for providing the tillite data, and Trond Slagstad for the chemical data in Grenville Province. We appreciate statistical insights from Steve Dye, Eligio Lisi, Michael Evans and Gerti Xhixha. Fruitful and insightful discussions with Vedran Lekic and Mark Chen improve our work. This study was funded U.S. National Science Foundation Grants EAR 1067983/1068097 and partially by the National Institute of Nuclear Physics (INFN), Italy.

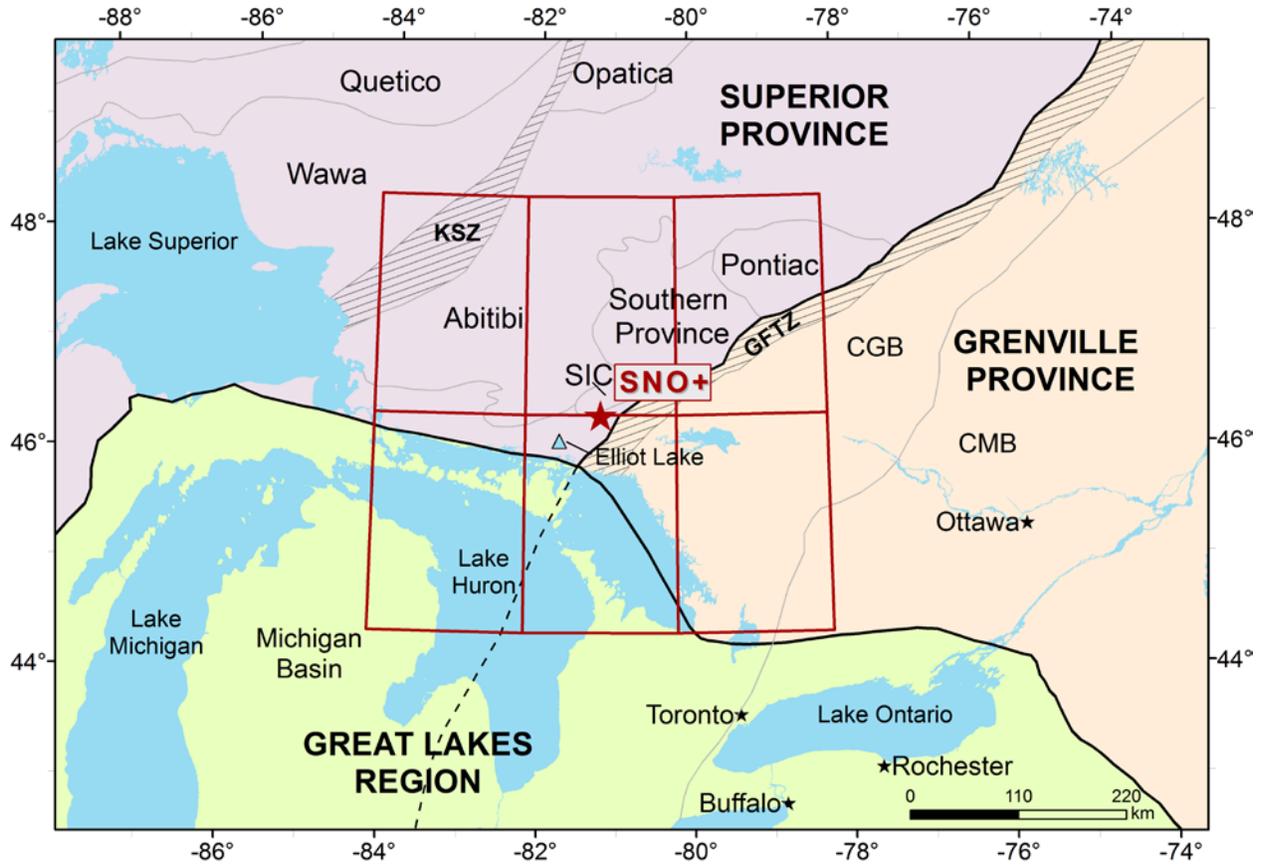

(a)

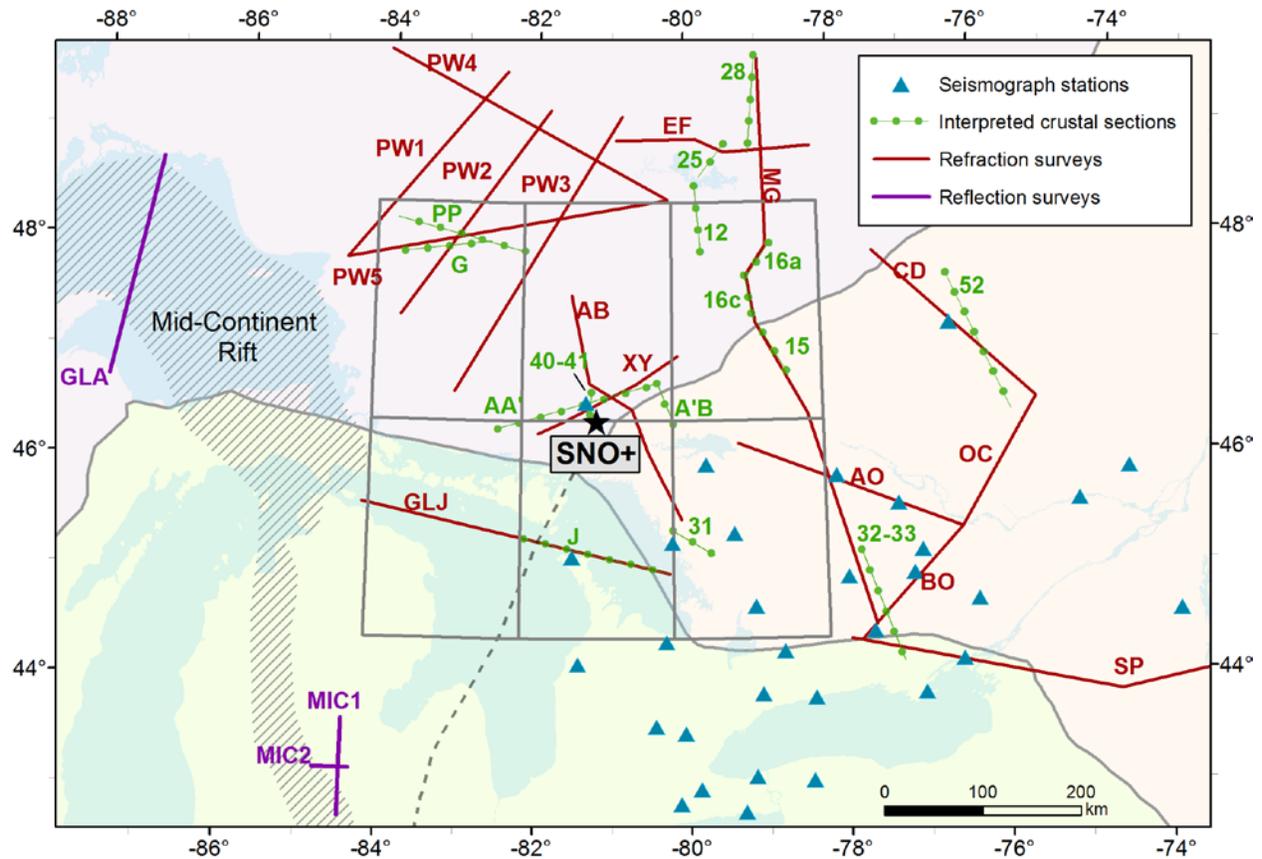

**(b)**

Fig 1: (a) Schematic map of the region around Sudbury, where the SNO+ (red star) is located. The six 2°×2° tiles (outlined by red lines) centered at SNO+ are chosen as the site of the 3-D regional crustal model. Light black lines show the major terrane boundaries within the Superior Province, which is separated from the Grenville Province by the Grenville Front Tectonic Zone (GFTZ). The Central Gneiss Belt (CGB) and Central Metasedimentary Belt (CMB) are two sub-terranes in the Grenville Province. The Great Lakes region is covered by Paleozoic sediments, extending to the Great Lakes Tectonic Zone (GLTZ) in the north. The Kapuskasing Structural Zone (KSZ) is a region of tectonically uplifted deep crustal rocks in the Superior Province. (b) Geophysical constraints in the 3-D crustal model are from local refraction surveys (red lines), reflection surveys (purple lines), interpreted crustal cross sections (green dotted lines) and broadband seismometers (blue triangles).

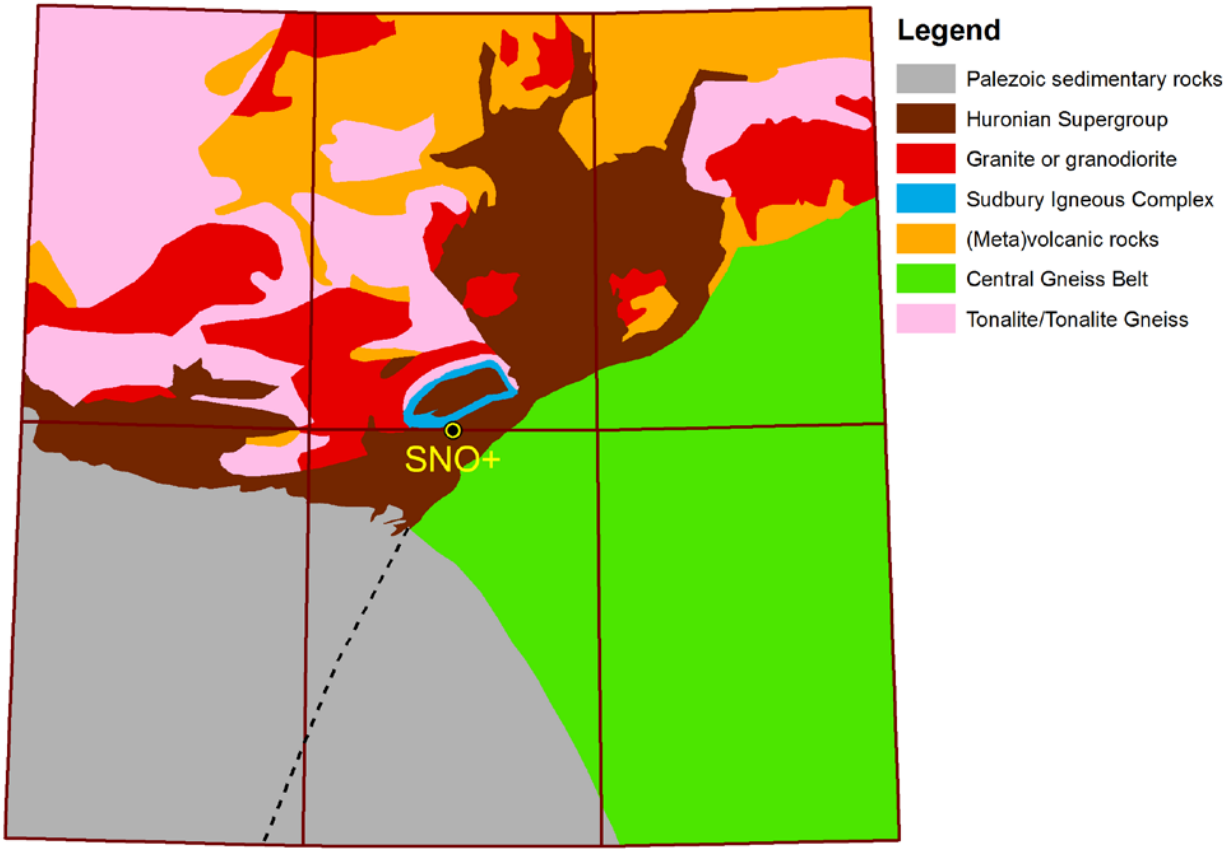

Fig. 2: Simplified geological map of the six tiles centered at SNO+. The upper crust is divided into seven dominant lithologic units that are assumed to be chemically uniform. The dashed black line is the extension of GFTZ (Grenville Front Tectonic Zone) under the Paleozoic sedimentary cover in the Great Lakes region. Color legends: pink – tonalite/tonalite gneiss in the Wawa and Abitibi sub-provinces; green – gneissic rocks in the Central Gneiss Belt (CGB) of the Grenville Province; orange – (meta)volcanic rocks in the Abitibi greenstone belt; gray – Paleozoic sedimentary cover in the Great Lakes; red – granitic to granodioritic intrusion in Wawa-Abitibi; brown – Huronian Supergroup sedimentary rocks and the sedimentary rocks of the Sudbury Basin; blue – the Sudbury Igneous Complex (SIC).

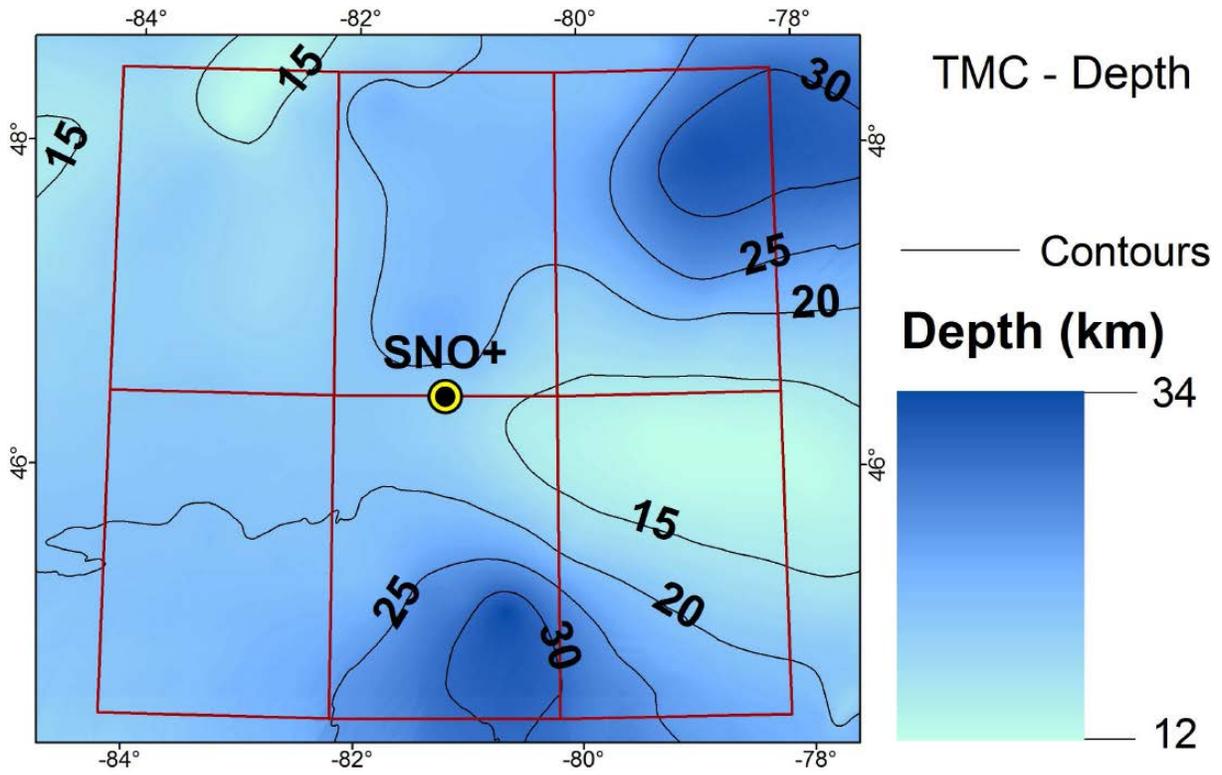

(a)

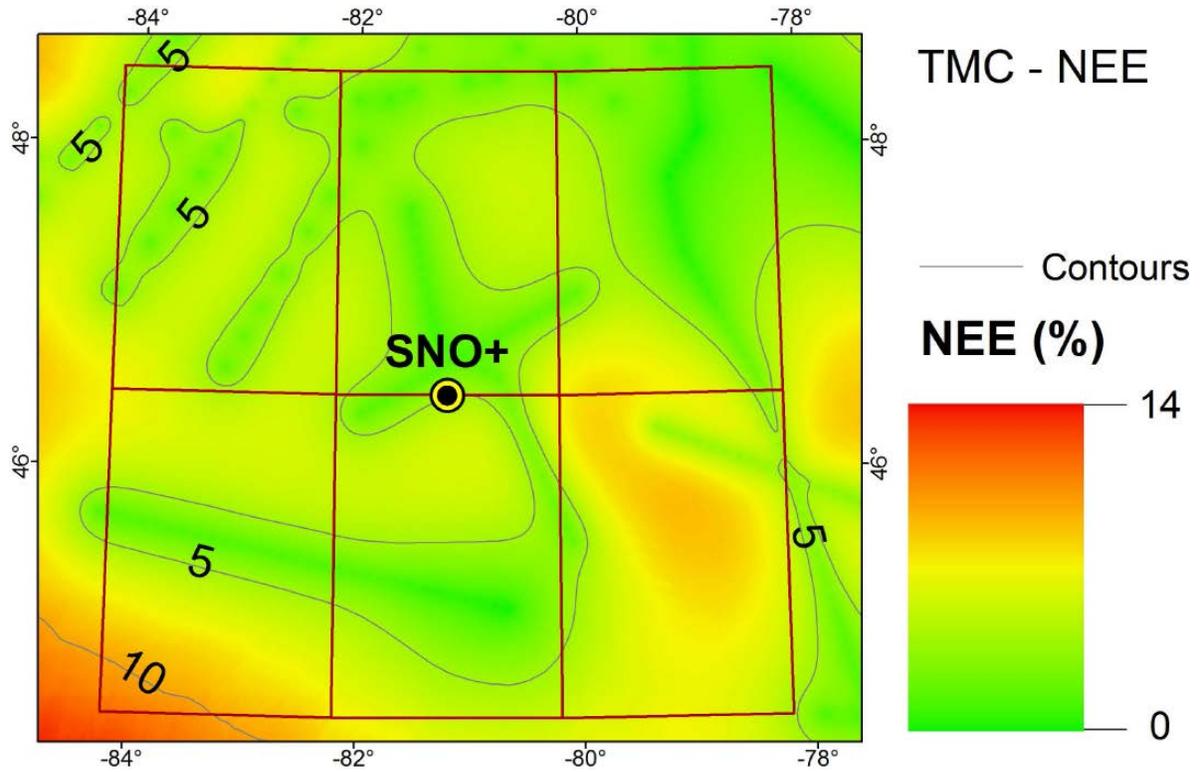

(b)

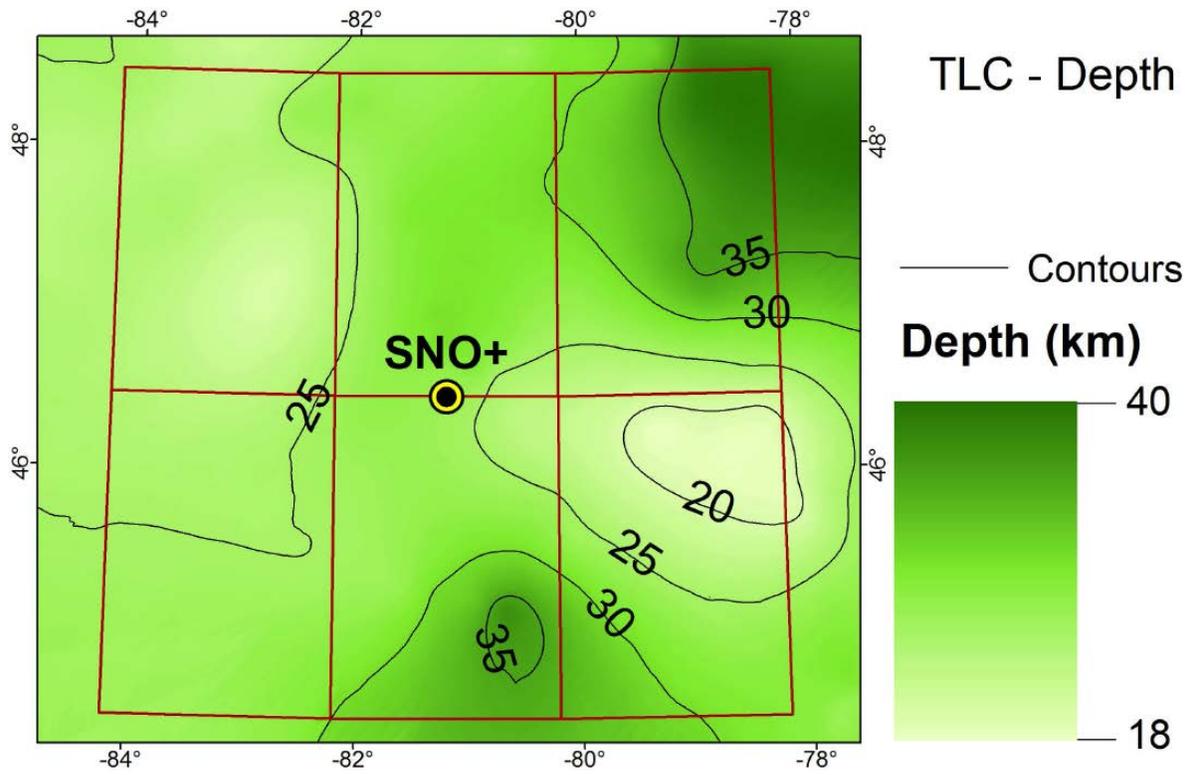

(c)

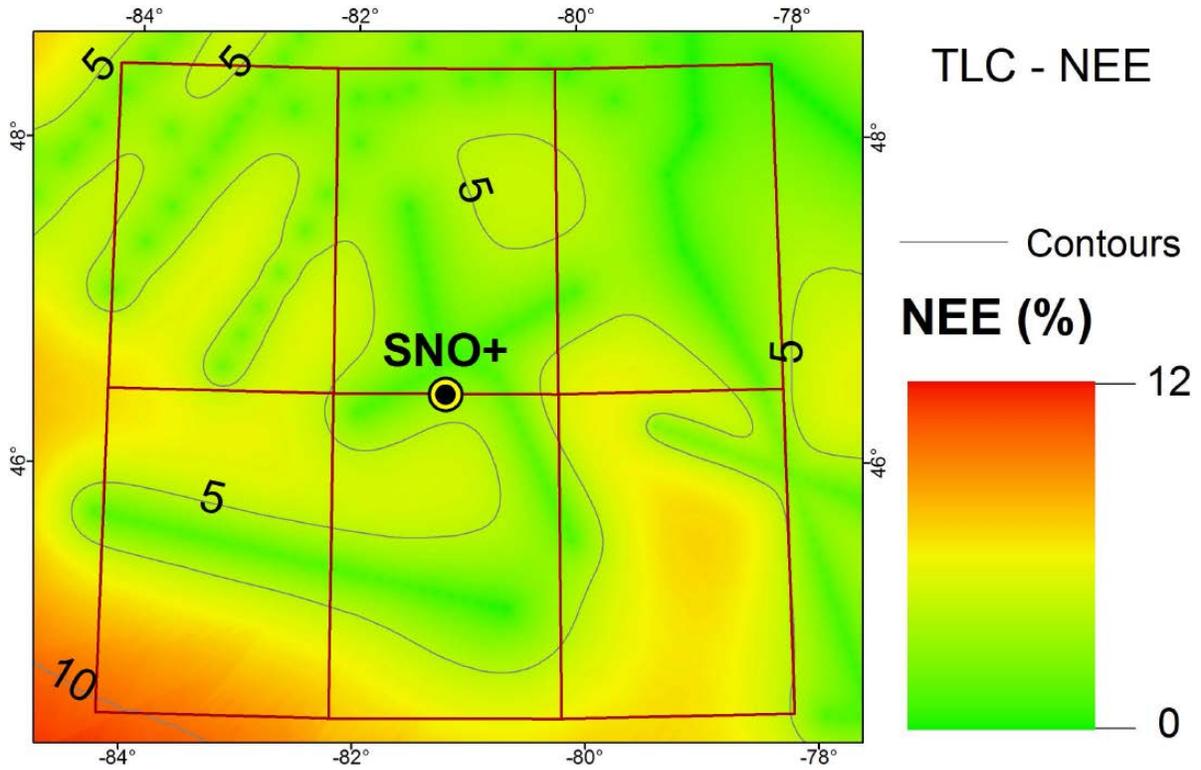

(d)

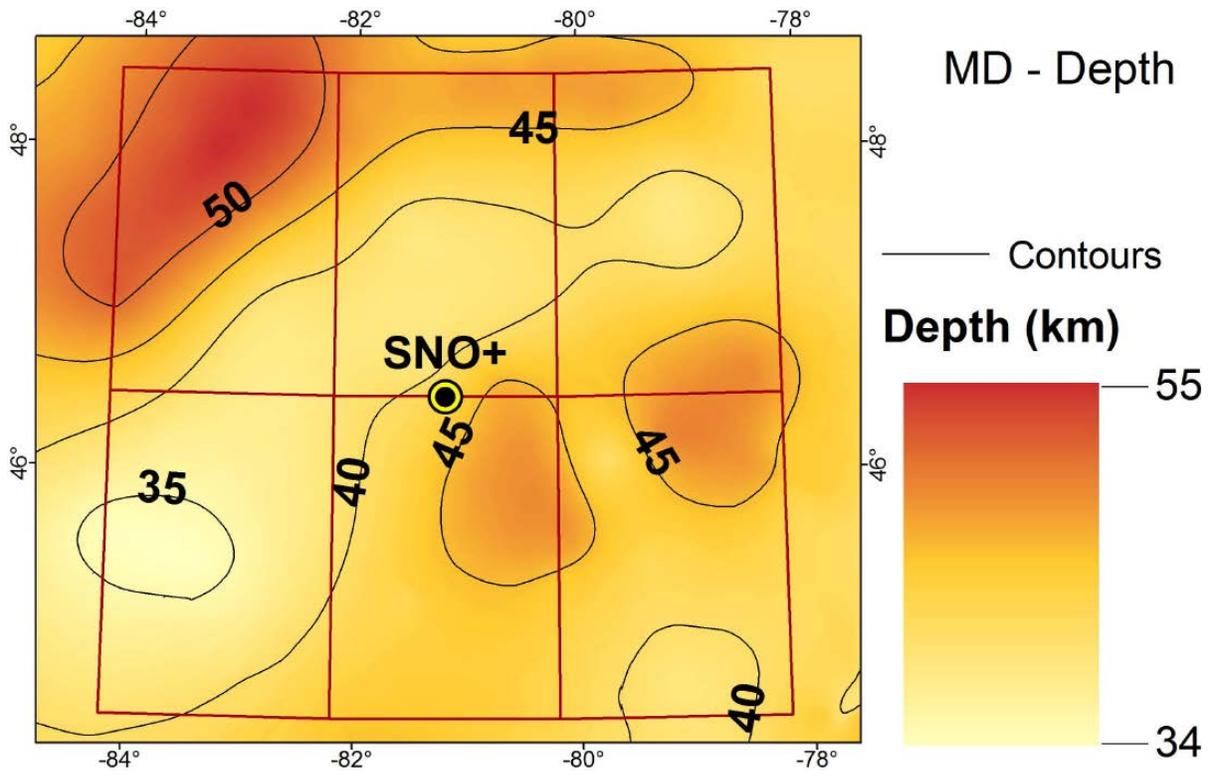

(e)

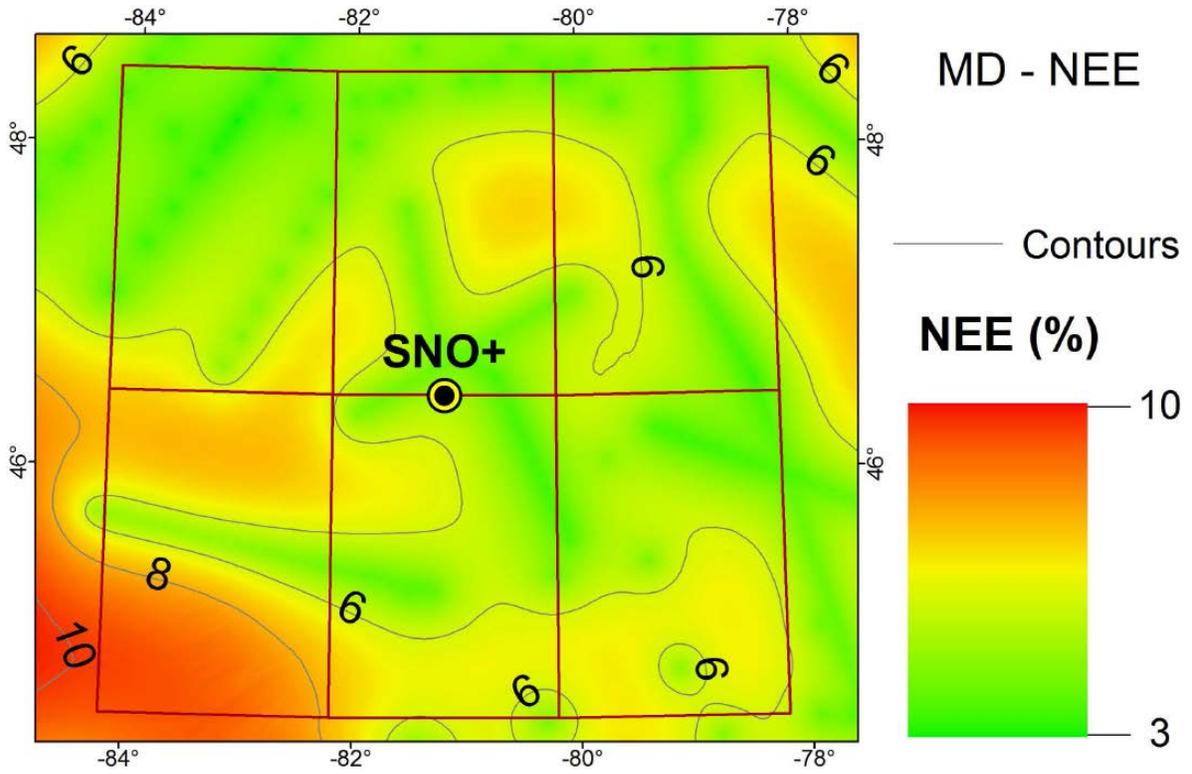

(f)

Fig. 3: Maps (1 km × 1 km resolution) of the depth of intracrustal boundaries in the 3-D regional crust model: (a, c, e) depth (km) of TMC (top of middle crust), TLC (top of lower crust) and MD (Moho depth), respectively; (b, d, f) normalized estimation errors (%) (NEE) for the estimated depth of TMC, TLC and MD, respectively.

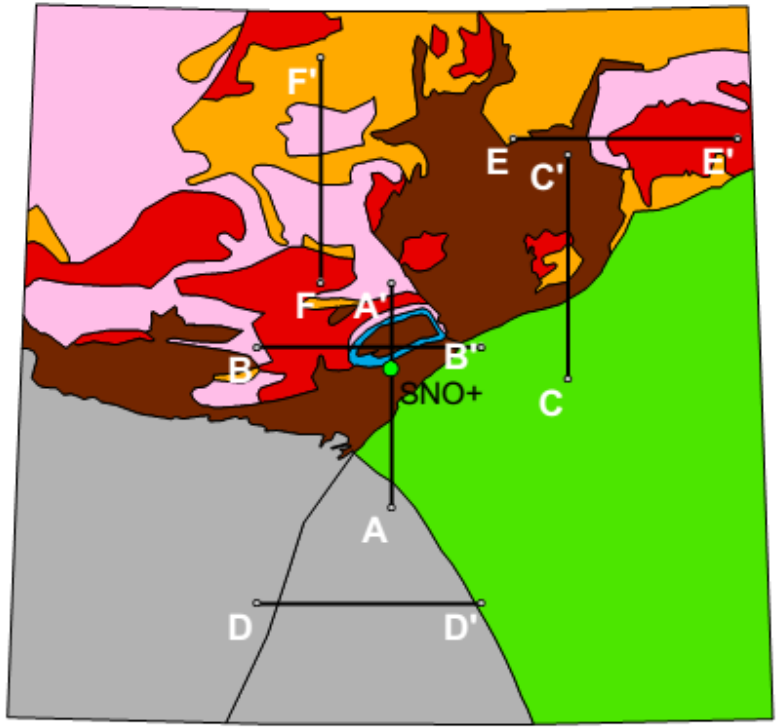

(a)

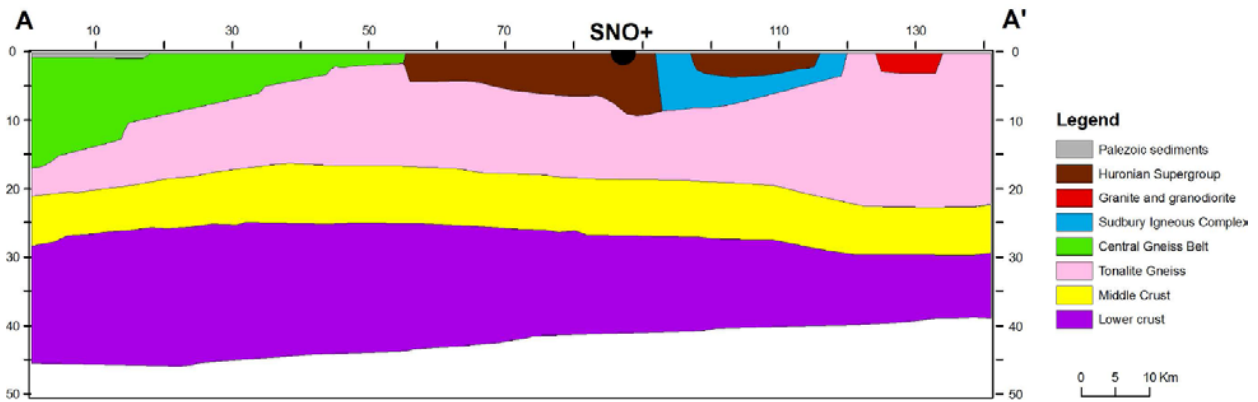

(b)

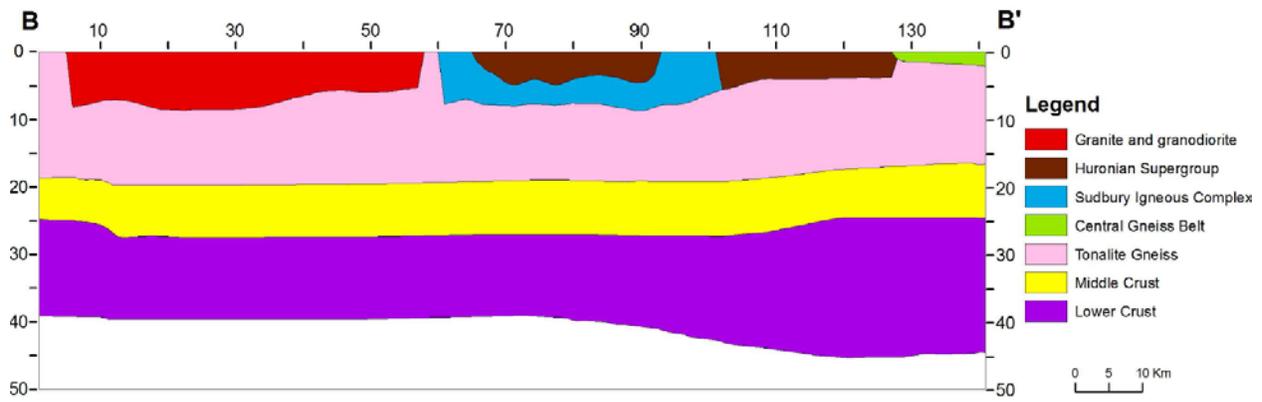

(c)

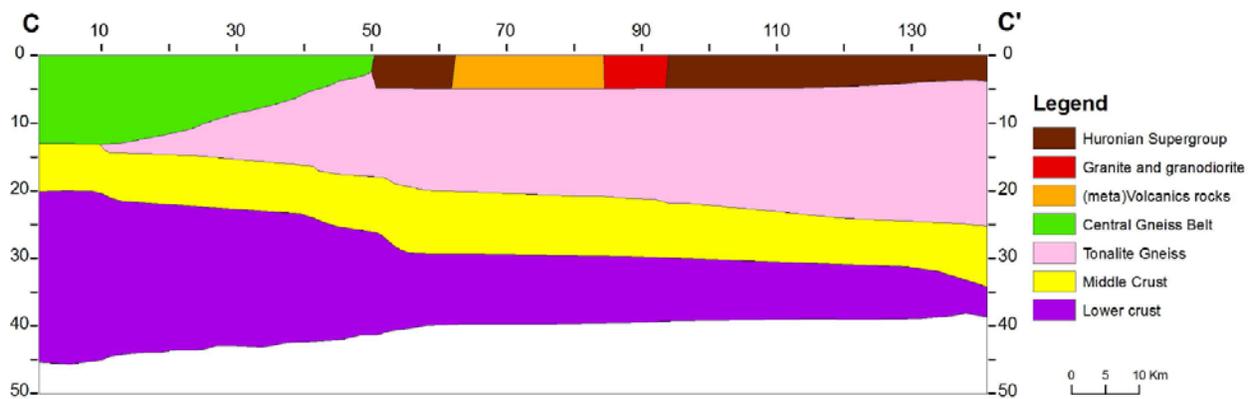

(d)

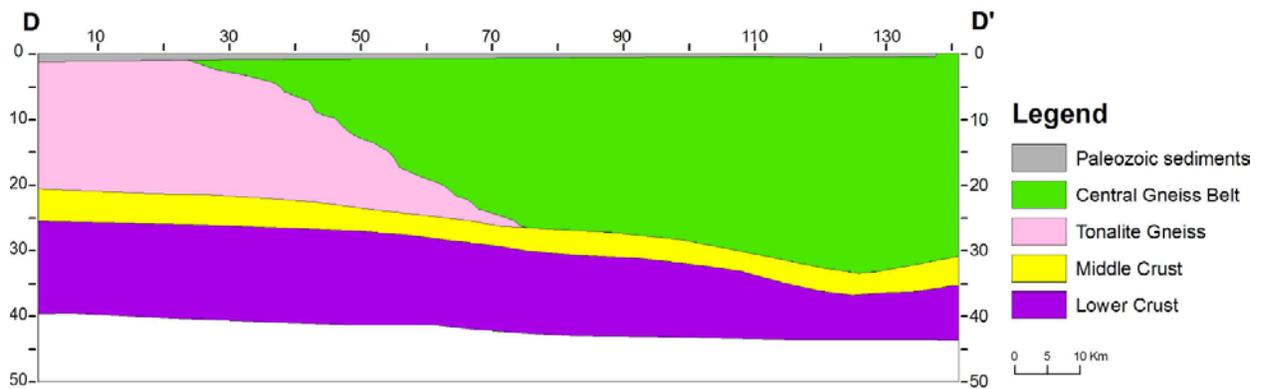

(e)

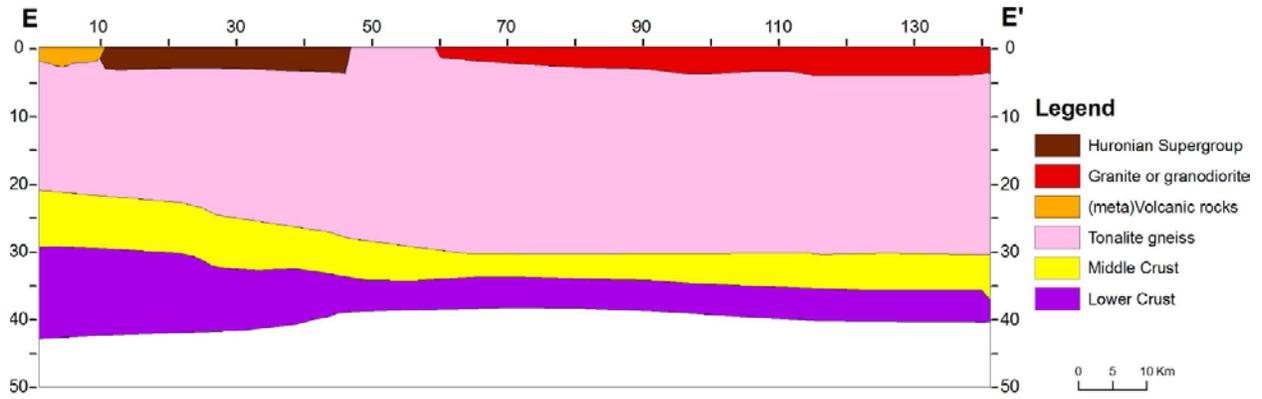

(f)

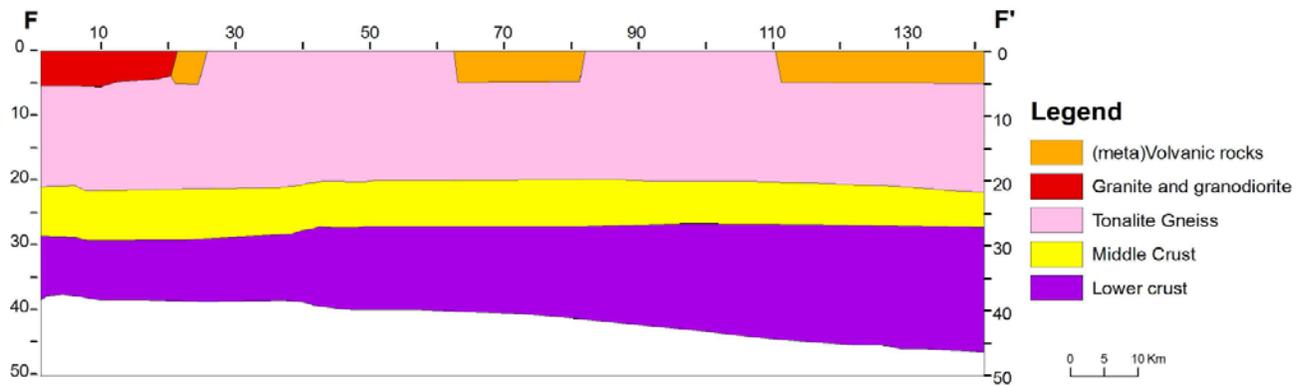

(g)

Fig. 4: Six schematic E-W and N-S cross sections shoing inferred crustal structure around SNO+ based on the 3-D model.

**Table 1: The five seismic experiments that provide the 18 seismic lines used to model the crust in the region surrounding SNO+.**

| Experiment | Main investigated area | Number of lines | Type[a] | Labels in Fig. 1B | Reference |
|---|---|---|---|---|---|
| LITHOPROBE | Sudbury Basin | 2 | RF | XY; AB | *Winardhi and Mereu* [1997] |
| | Superior Province | 2 | RF | EF; MG | *Winardhi and Mereu* [1997] |
| | Kapuskasing Structural Zone | 5 | RF | PW1; PW2; PW3; PW4; PW5 | *Percival and West* [1994] |
| COCRUST | Grenville Province | 4 | RF | AO; OB; BC; CD | *Mereu et al.* [1986] |
| O-NYNEX | Appalachian Province | 1 | RF | SP | *Musacchio et al.* [1997] |
| GLIMPCE | Great Lakes | 1 | RF | GLJ | *Epili and Mereu* [1991] |
| | | 1 | RL | GLA | *Spence et al.* [2010] |
| COCORP | Michigan Basin | 2 | RL | MIC1; MIC2 | *Brown et al.* [1982] |

[a]RF: refraction; RL: reflection

**Table 2: The interpreted crustal cross sections used to define the contact between the seven dominant lithologic unit in the physical model of upper crust in the region surrounding SNO+.**

| Main investigated areas | Labels in Fig. 1b | Reference |
|---|---|---|
| Sudbury Basin | AA' - A'B | *Easton* [2000] |
| | 40 - 41 | *Adam et al.* [2000] |
| Superior Province | 12 - 15 - 16a - 16c - 25 - 28 | *Ludden and Hynes* [2000] |
| Kapuskasing Structural Zone | PP | *Percival and Peterman* [1994] |
| | G | *Geis et al.* [1990] |
| Grenville Province | 31 - 32 - 33 | *White et al.* [2000] |
| Great Lakes | J | *White et al.* [2000] |

**Table 3:** Descriptive statistics of depth-controlling data points from refraction and reflection seismic profiles and receiver function analyses that are used to produce the TMC, TLC and MD depth maps. The minimum, mean and maximum values of the data points along each boundary surface are reported.

|  | N° point input | Variance (km$^2$) | Depth (km) | | |
|---|---|---|---|---|---|
|  |  |  | Min | Mean | Max |
| **TMC** | 343 | 24.5 | 10.0 | 19.7 | 35.0 |
| **TLC** | 343 | 27.6 | 18.0 | 28.3 | 40.0 |
| **MD** | 397 | 23.2 | 31.0 | 42.6 | 56.0 |

**Table 4: Physical properties (thickness, volume, volume fraction, density) of upper, middle and lower crust within the 3-D regional model, and comparisons with previous models.**

|       | Thickness (km) | | | Volume ($10^6$ km$^3$) | Volume (%) | $\rho$ (g/cm$^3$) | | |
|-------|----------------|---|---|------------------------|------------|-------------------|---|---|
|       | C 1.0[a]       | H'13[b] | This study | | | C 1.0 | H'13 | This study |
| UC    | 11.8 | 14.7 ± 1.0 | 20.3 ± 1.1 | 4.2 ± 0.2 | 47.9 | 2.75 | 2.80 | 2.73 ± 0.08 |
| MC    | 14.1 | 15.2 ± 1.0 | 6.4 ± 0.4  | 1.3 ± 0.1 | 15.3 | 2.82 | 2.88 | 2.96 ± 0.03 |
| LC    | 13.3 | 12.7 ± 0.8 | 15.6 ± 1.0 | 3.2 ± 0.2 | 36.8 | 2.92 | 3.03 | 3.08 ± 0.06 |
| Total | 39.2 | 42.6 ± 2.8 | 42.3 ± 2.6 | 8.7 ± 0.5 | 100  | ==   | ==   | == |

[a]C 1.0 is the updated CRUST1.0 model by *Laske et al.* [2013] at http://igppweb.ucsd.edu/~gabi/rem.html;
[a]No uncertainty is provided in CRUST 1.0 for the thickness of crustal layers;
[b]Average of individual uncertainties for the 24 1° × 1° voxels in the reference model of *Huang et al.* [2013].

**Table 5: Physical properties (geology feature, thickness, volume, volume fraction, density and mass) of the seven dominant lithologic units in the upper crust within the 3-D regional model.**

| Lithology/Geology | Average Thickness (km) | Volume ($10^6$ km$^3$) | Volume (%) | $\rho$ (g/cm$^3$) | Mass ($10^{18}$ kg) |
|---|---|---|---|---|---|
| Tonalite/Tonalite gneiss (Wawa-Abitibi) | 16.6 | 2.51 | 60.6 | $2.73 \pm 0.08^a$ | $6.9 \pm 0.5$ |
| Central Gneiss Belt (Grenville Province) | 14.5 | 1.25 | 30.2 | $2.73 \pm 0.08^b$ | $3.4 \pm 0.2$ |
| (Meta)volcanic rocks (Abitibi sub-province) | 5.5 | 0.12 | 2.9 | $2.84 \pm 0.14^a$ | $0.34 \pm 0.01$ |
| Paleozoic sedimentary rocks | 1.1 | 0.06 | 1.3 | $2.62 \pm 0.19^c$ | $0.16 \pm 0.01$ |
| Granite or granodiorite (Wawa-Abitibi) | 5.2 | 0.09 | 2.2 | $2.67 \pm 0.02^a$ | $0.24 \pm 0.00$ |
| Huronian Supergroup, Sudbury Basin | 4.4 | 0.11 | 2.7 | $2.69 \pm 0.04^d$ | $0.30 \pm 0.01$ |
| Sudbury Igneous Complex | 6.1 | 0.006 | 0.1 | $2.8 \pm 0.1^e$ | $0.02 \pm 0.00$ |

$^a$Average values of data from *Fountain & Salisbury* [1996], *Fountain et al.* [1990] and *Salisbury & Fountain* [1994].
$^b$Density of green unit is assumed equal to the density of the pink unit.
$^c$Average values from data taken by *Hinze et al.* [1978].
$^d$Ontario Geological Survey - Geophysical Series - Preliminary map. 2297 - North Bay Marten River Area – Districts of Nipissing and Sudbury. Scale 1:100.000 (1980).
$^e$*Snyder et al.* [2002], *Milkereit et al.* [1994].

**Table 6: U and Th abundances in seven dominant lithologic units in the regional upper crust near SNO+ in the 3-D model.**

| Lithologic Unit | | | U mean[a] | 1-sigma + | 1-sigma - | median | n | Th mean[a] | 1-sigma + | 1-sigma - | median | n | Correlation[b] |
|---|---|---|---|---|---|---|---|---|---|---|---|---|---|
| Tonalite/Tonalite gneiss | | all | 0.7 | 1.0 | 0.4 | 0.7 | 141 | 3.0 | 4.6 | 1.8 | 3.2 | 146 | 0.74 |
| | | filtered[c] | **0.7** | **0.5** | **0.3** | 0.7 | 111 | **3.1** | **2.3** | **1.3** | 3.1 | 107 | |
| Gneiss in CGB | | all | 2.6 | 0.4 | 0.4 | 2.7 | 5 | 3.9 | 8.9 | 2.7 | 5.3 | 96 | -- |
| | | filtered | **2.6** | **0.4** | **0.4** | 2.7 | 5 | **5.1** | **6.0** | **2.8** | 5.9 | 68 | |
| (Meta)volcanic rocks | Felsic (5%) | all | 1.1 | 1.7 | 0.7 | 1.0 | 472 | 4.3 | 6.7 | 2.6 | 4.3 | 531 | 0.86 |
| | | filtered | **1.1** | **0.8** | **0.5** | 1.0 | 402 | **4.3** | **3.0** | **1.8** | 4.1 | 416 | |
| | Intermediate (40%) | all | 0.5 | 1.1 | 0.3 | 0.5 | 192 | 1.6 | 3.3 | 1.1 | 1.6 | 246 | 0.87 |
| | | filtered | **0.5** | **0.4** | **0.2** | 0.5 | 135 | **1.5** | **1.3** | **0.7** | 1.6 | 170 | |
| | Mafic (55%) | all | 0.3 | 0.8 | 0.2 | 0.3 | 333 | 0.9 | 2.4 | 0.6 | 0.8 | 414 | 0.88 |
| | | filtered | **0.2** | **0.4** | **0.1** | 0.2 | 249 | **0.8** | **1.0** | **0.4** | 0.7 | 316 | |
| Paleozoic sedimentary rocks | | all | 3.1 | 5.5 | 2.0 | 2.5 | 10606 | 4.5 | 3.0 | 1.8 | 4.4 | 2196 | 0.55 |
| | | filtered | **2.5** | **2.0** | **1.1** | 2.3 | 8466 | **4.4** | **1.6** | **1.2** | 4.3 | 1700 | |
| Felsic intrusion | Granite (60%) | all | 3.9 | 4.1 | 2.0 | 4.1 | 26 | 24.1 | 26.8 | 12.7 | 28.0 | 25 | 0.60 |
| | | filtered | **4.0** | **2.3** | **1.4** | 4.1 | 18 | **29.7** | **12.0** | **8.6** | 28.9 | 19 | |
| | Granodiorite (40%) | all | 1.1 | 0.8 | 0.5 | 1.1 | 92 | 5.4 | 6.2 | 2.9 | 5.5 | 92 | 0.81 |
| | | filtered | **1.2** | **0.5** | **0.3** | 1.2 | 70 | **5.2** | **3.1** | **2.0** | 5.2 | 69 | |
| Huronian Supergroup, Sudbury Basin | | all | 4.2 | 6.4 | 2.5 | 4.1 | 207 | 11.8 | 20.8 | 7.5 | 11.4 | 214 | 0.90 |
| | | filtered | **4.2** | **2.9** | **1.7** | 4.2 | 156 | **11.1** | **8.2** | **4.8** | 11.3 | 177 | |
| Sudbury Igneous Complex | Norite (40%) | all | 1.1 | 0.5 | 0.3 | 1.2 | 80 | 5.6 | 1.6 | 1.2 | 5.7 | 80 | 0.76 |
| | | filtered | **1.2** | **0.2** | **0.2** | 1.3 | 71 | **5.7** | **0.7** | **0.7** | 5.7 | 72 | |
| | Quartz Gabbro (10%) | all | 1.7 | 0.5 | 0.4 | 1.6 | 19 | 7.5 | 2.4 | 1.8 | 6.7 | 19 | 0.99 |
| | | filtered | **1.5** | **0.2** | **0.2** | 1.5 | 13 | **6.7** | **0.9** | **0.8** | 6.6 | 14 | |
| | Granophyre (50%) | all | 3.3 | 0.2 | 0.2 | 3.2 | 25 | 14.9 | 1.0 | 1.0 | 14.8 | 25 | 0.95 |
| | | filtered | **3.3** | **0.1** | **0.1** | 3.2 | 18 | **15.2** | **0.7** | **0.6** | 15.3 | 18 | |

[a] Log-normal mean, see *Huang et al.* [2013].
[b] Correlation between logarithmic values of the U and Th abundances
[c] A filter was applied to remove 25% of the raw data in order to reduce the influence of rare enriched or depleted samples on the log-normal mean

**Table 7: Abundances of U and Th (ppm), heat production (μW m⁻³) and geoneutrino signal (TNU) at SNO+ from all reservoirs in the 3-D regional model.**

| Lithology/Geology | A(U) | A(Th) | H(U) | H(Th) | S(U) | S(Th) | S(U+Th) |
|---|---|---|---|---|---|---|---|
| Tonalite/Tonalite gneiss (Wawa-Abitibi) | $0.7^{+0.5}_{-0.3}$ | $3.1^{+2.3}_{-1.3}$ | $0.2^{+0.1}_{-0.1}$ | $0.22^{+0.17}_{-0.09}$ | $1.6^{+1.1}_{-0.7}$ | $0.5^{+0.4}_{-0.2}$ | $2.2^{+1.4}_{-0.9}$ |
| Central Gneiss Belt (Grenville Province) | $2.6^{+0.4}_{-0.4}$ | $5.1^{+6.0}_{-2.8}$ | $0.70^{+0.11}_{-0.10}$ | $0.36^{+0.43}_{-0.20}$ | $1.8^{+0.3}_{-0.3}$ | $0.2^{+0.3}_{-0.1}$ | $2.1^{+0.4}_{-0.3}$ |
| (Meta)volcanic rocks (Abitibi sub-province) | $0.4^{+0.3}_{-0.1}$ | $1.5^{+0.9}_{-0.5}$ | $0.1^{+0.1}_{-0.0}$ | $0.11^{+0.06}_{-0.04}$ | $0.02^{+0.01}_{-0.01}$ | $0.004^{+0.003}_{-0.002}$ | $0.02^{+0.01}_{-0.01}$ |
| Paleozoic sediments | $2.5^{+2.0}_{-1.1}$ | $4.4^{+1.6}_{-1.2}$ | $0.65^{+0.51}_{-0.29}$ | $0.30^{+0.11}_{-0.08}$ | $0.04^{+0.03}_{-0.02}$ | $0.01^{+0.00}_{-0.00}$ | $0.05^{+0.04}_{-0.02}$ |
| Granite or granodiorite (Wawa-Abitibi) | $2.8^{+1.4}_{-0.8}$ | $19.6^{+7.1}_{-5.2}$ | $0.76^{+0.36}_{-0.23}$ | $1.41^{+0.51}_{-0.37}$ | $0.3^{+0.2}_{-0.1}$ | $0.2^{+0.1}_{-0.0}$ | $0.5^{+0.2}_{-0.1}$ |
| Huronian Supergroup, Sudbury Basin | $4.2^{+2.9}_{-1.7}$ | $11.1^{+8.2}_{-4.8}$ | $1.13^{+0.77}_{-0.46}$ | $0.78^{+0.59}_{-0.33}$ | $6.2^{+4.3}_{-2.6}$ | $1.1^{+0.8}_{-0.5}$ | $7.3^{+5.0}_{-3.0}$ |
| Sudbury Igneous Complex | $2.3^{+0.2}_{-0.1}$ | $10.7^{+0.4}_{-0.4}$ | $0.63^{+0.03}_{-0.02}$ | $0.77^{+0.03}_{-0.03}$ | $0.6^{+0.0}_{-0.0}$ | $0.2^{+0.0}_{-0.0}$ | $0.8^{+0.0}_{-0.0}$ |
| Upper crust | $1.4^{+0.3}_{-0.2}$ | $4.6^{+2.3}_{-1.4}$ | $0.38^{+0.08}_{-0.06}$ | $0.33^{+0.17}_{-0.10}$ | $10.9^{+4.4}_{-2.8}$ | $2.4^{+1.0}_{-0.6}$ | $13.3^{+5.2}_{-3.3}$ |
| Middle crust | $0.8^{+0.5}_{-0.3}$ | $3.5^{+2.8}_{-1.6}$ | $0.2^{+0.1}_{-0.1}$ | $0.25^{+0.20}_{-0.11}$ | $0.9^{+0.5}_{-0.3}$ | $0.3^{+0.2}_{-0.1}$ | $1.2^{+0.7}_{-0.4}$ |
| Lower crust | $0.2^{+0.2}_{-0.1}$ | $1.4^{+1.8}_{-0.7}$ | $0.06^{+0.04}_{-0.02}$ | $0.10^{+0.13}_{-0.05}$ | $0.5^{+0.3}_{-0.2}$ | $0.2^{+0.3}_{-0.1}$ | $0.7^{+0.6}_{-0.3}$ |
| Total Crust | $0.9^{+0.2}_{-0.1}$ | $3.6^{+0.6}_{-0.7}$ | $0.3^{+0.1}_{-0.0}$ | $0.26^{+0.11}_{-0.07}$ | $12.5^{+4.4}_{-2.8}$ | $3.0^{+1.0}_{-0.7}$ | $15.6^{+5.3}_{-3.4}$ |